\documentclass[a4paper,11pt]{article}
\usepackage{jheppub} 
\usepackage{lineno}
\usepackage{amsmath}
\usepackage{physics}
\usepackage{subfigure}

\def\bc{\begin{center}}

\def\ec{\end{center}}
\def\be{\begin{eqnarray}}
\def\ee{\end{eqnarray}}

\title{\boldmath Probing Krylov Complexity in Scalar Field Theory with General Temperatures}

\author[a]{Peng-Zhang He}
\author[a,b]{, Hai-Qing Zhang}
\affiliation[a]{Center for Gravitational Physics, Department of Space Science, Beihang University, Beijing 100191, China}
\affiliation[b]{Peng Huanwu Collaborative Center for Research and Education, Beihang University, Beijing 100191, China}

\emailAdd{hepzh@buaa.edu.cn}
\emailAdd{hqzhang@buaa.edu.cn}

\abstract{Krylov complexity characterizes the operator growth in the quantum many-body systems or quantum field theories. The existing literatures have studied the Krylov complexity in the low temperature limit in the quantum field theories. In this paper, we extend and systematically study the Krylov complexity and Krylov entropy in a scalar field theory with general temperatures. To this end, we propose a new method to calculate the Wightman power spectrum which allows us to compute the Lanczos coefficients and subsequently to study the Krylov complexity (entropy) in general temperatures. We find that the Lanczos coefficients and Krylov complexity (entropy) in the high temperature limit will behave somewhat differently from those studies in the low temperature limit. We give an explanation of why the Krylov complexity does not oscillate in the high-temperature region. Moreover, we uncover the transition temperature that separates the oscillating and monotonic increasing behavior of Krylov complexity.}

\begin{document}
\maketitle
\flushbottom
    \section{Introduction}
    In recent years, complexity \cite{Nielsen:2006cea,Jefferson:2017sdb,Parker:2018yvk} has become a crucial notion to understand the chaotic properties of a quantum system or spacetime \cite{PhysRevLett.116.191301,Susskind:2014rva}. It provides a new insight and an analytical approach to probe the chaotic behaviors of a physical system. The complexity is originated as a means to quantify the difficulty of a quantum system which can transit from one state to another. With the deepening of the research, complexity has received wide applications in the fields such as quantum chaos, quantum information, quantum field theory, and holographic theory, etc. \cite{Aaronson:2016vto}.
    
    Recent studies on complexity has led to the emergence of Krylov complexity as an important concept in quantum physics. Initially, it was developed to explore the growth of operator in the Heisenberg picture and to distinguish between chaotic and integrable systems \cite{Parker:2018yvk}. Usually, this operator is defined in a local Hilbert space called {\it Krylov space}, which is the span of nested commutators (see the following section \ref{sec2}). The {\it Lanczos coefficients} can be constructed according to the basis in the Krylov space. Consequently, Krylov complexity can be computed from the Schr\"odinger-type equation of the Lanczos coefficients.  The authors \cite{Parker:2018yvk} found that the time evolution of the Krylov complexity satisfy an exponential growth as time goes by, which indicates a signature of chaos in quantum many-body systems, such as the spin chains and Sachdev-Ye-Kitaev (SYK) model \cite{Sachdev:1992fk,kitaev}. At finite temperature, the authors further found that there is an upper bound for this exponential growth rate of the Krylov complexity. 
    
    In nowadays, Krylov complexity has received increasingly widespread applications in many other quantum systems \cite{Bhattacharya:2022gbz,Bhattacharjee:2022lzy,Liu:2022god,Hashimoto:2023swv,Adhikari:2022whf,Avdoshkin:2022xuw}. Specifically, the study of Krylov complexity has unveiled some astonishing phenomena. In chaotic systems, its growth initially exhibits an exponential trend, gradually transits to linear, and eventually reaches a plateau, indicating a saturation, as the system size is finite \cite{Parker:2018yvk,Rabinovici:2020ryf}. However, even in integrable quantum systems, such as free scalar field theory \cite{Camargo:2022rnt} or conformal field theory \cite{Dymarsky:2021bjq}, the Lanczos coefficients display linear growth, while the Krylov complexity exhibits an exponential growth. The paper \cite{Bhattacharjee:2022vlt} is the first to present non-trivial evidence that integrable systems, under specific conditions, can exhibit the linear behavior of the Lanczos coefficients. This implies that there are still some subtle aspects of Krylov complexity that we have not yet fully understood. So far, the study of Krylov complexity has permeated multiple fields such as free and interacting field theory, random matrix theory, and open quantum systems etc \cite{Bhattacharjee:2022qjw,Balasubramanian:2023kwd,Dymarsky:2019elm,Dymarsky:2021bjq,Camargo:2022rnt,Vasli:2023syq,PhysRevD.106.126022,Iizuka:2023pov,Iizuka:2023fba,Erdmenger:2023wjg,Bhattacharjee:2022ave,Bhattacharjee:2023uwx,Bhattacharya:2023zqt,Bhattacharyya:2023grv,Malvimat:2024vhr,Caputa:2024vrn,Afrasiar:2022efk,Tan:2024kqb,Li:2024iji,Li:2024kfm,Vardian:2024fsp,Camargo:2023eev,Huh:2023jxt,Camargo:2024deu}. Moreover, the concept of Krylov complexity has been successfully applied to various specific problems, including the SYK model \cite{Rabinovici:2020ryf,He:2022ryk}, generalized coherent states \cite{Patramanis:2021lkx,Caputa:2021sib}, Ising and Heisenberg models \cite{Cao:2020zls,Trigueros:2021rwj,Heveling:2022hth}, topological phases of matter \cite{Caputa:2022eye}. Interested readers can refer to the recent seminal review \cite{Nandy:2024htc}.
    
    In this paper, we extend the study of Krylov complexity in five dimensional free field theory from \cite{Camargo:2022rnt} to general temperatures. In \cite{Camargo:2022rnt} the authors only studied the case in the low temperature limit which satisfies $\beta m\gg1$,\footnote{$\beta$ is the inverse of the temperature while $m$ is the mass of the field. Their definitions are defined in the following context. } which can simplify the Wightman power spectrum in this limit. However, in our paper we find a new method to compute the Wightman power spectrum with general $\beta m$, i.e., with general temperatures. As a result, the Lanczos coefficients can be subsequently computed and the Krylov complexity and Krylov entropy can also be obtained accordingly. We found that the Lanczos coefficients $b_n$ exhibit the `staggering' behavior, which separates the odd and even $n$ into two families. For large $n$ one can linearly fit the Lanczos coefficients as $b_n\sim\alpha n+\gamma$. The linear coefficient $\alpha$ is approximately identical to $\pi/\beta$. The differences $\Delta\gamma$ between $\gamma$ of odd $n$ and even $n$ is linearly proportional to the mass of the field as $\beta m\gg1$. These behaviors are consistent with previous studies in \cite{Camargo:2022rnt}.  However, in the high temperature limit $\beta m\ll1$, we observe that the linear behavior of the difference $\Delta\gamma$ to the mass will break down, which may come from the effect of the high temperature. We also found that in the high temperature limit, the Krylov complexity and Krylov entropy behaves like those in conformal field theories \cite{Dymarsky:2021bjq}. The exponential growth rate of the Krylov complexity in the high temperature limit is roughly $2\pi/\beta$, however, in the low temperature limit the rate is smaller than $2\pi/\beta$. Therefore, totally for the general temperatures, the exponential growth rate of the Krylov complexity has an upper bound $2\pi/\beta$ which is consistent with previous reports. For the Krylov entropy, we found that in the late time it is linearly proportional to time. This is consistent with the claims that Krylov entropy is proportional to logarithmic of the Krylov complexity. Interestingly, we found that the Krylov complexity in high-temperature limit does not oscillate, which has sharp comparison to those in low-temperature limit. We postulate that the different behaviors of Krylov complexity in high and low temperature limit are due to the different behavior of $b_n$ as well as the auto-correlation function $\varphi_0(t)$. Moreover, we numerically uncover the transition temperature at around $\beta m=10$ which separates the oscillations and monotonic increasing of the Krylov complexity. 

    This paper is arranged as follows: In section \ref{sec2} we will give a brief review of Krylov space and Krylov complexity. In Section \ref{sec3}, we will briefly review some of the results from \cite{Camargo:2022rnt} and provide a physical explanation for the oscillation of the Krylov complexity in the low-temperature region;  In section \ref{sec4}, we will first introduce a new method to compute the Wightman power spectrum in general temperatures, and then compare the Krylov complexity and Krylov entropy for general temperatures in the free scalar field theory; Then we draw our conclusions in section \ref{sec5}. In the appendix \ref{secssb}, we will discuss the Krylov complexity and Krylov entropy in the spontaneous symmetry breaking phase. Appendix \ref{A} will provide some details about the spontaneous symmetry breaking in the real scalar field theory. These two appendices can be regarded as an application of Krylov complexity in high-temperature from section \ref{sec4.1}.

	\section{Overview on Krylov Space and Krylov Complexity}\label{sec2}
In this section, we will give a brief review on the basics of Krylov space and Krylov complexity. Interested readers can refer to Refs. \cite{Parker:2018yvk,Camargo:2022rnt,Nandy:2024htc} for more details. 
\subsection{Lanczos algorithm}
In quantum mechanics, time evolution of an operator $\mathcal{O}$ is determined by the Heisenberg equation
	\begin{equation}\label{1.1}
		\partial_{t}\mathcal{O}(t)=i[H,\mathcal{O}(t)],
	\end{equation}
	where $H$ is the Hamiltonian of the system. The solution of the above equation is
	\begin{equation}\label{1.22}
		\mathcal{O}(t)=e^{iHt}\mathcal{O}(0)e^{-iHt}.
	\end{equation} 
	 where $\mathcal{O}(0)$ is the value of the operator at time $t=0$. From the well-known Baker-Campbell-Hausdorff formula \cite{wachter2011relativistic}
	\begin{equation}
		e^{A}Be^{-A}=\sum_{n=0}^{\infty}\frac{1}{n!}[A^{(n)},B],
	\end{equation}
    where $[A^{(n)},B]\equiv[\underbrace{A,\cdots,[A,[A}_n, B]]$, the Eq.\eqref{1.22} can be rewritten as
	\begin{equation}
		\mathcal{O}(t)=\sum_{n=0}^{\infty}\frac{(it)^{n}}{n!}[H^{(n)},\mathcal{O}]\equiv \sum_{n=0}^{\infty}\frac{(it)^{n}}{n!}\tilde{\mathcal{O}}_{n},
	\end{equation}
 where we have defined  $\tilde{\mathcal{O}}_{n}\equiv[H^{(n)},\mathcal{O}]$ and set $\mathcal{O}\equiv\mathcal{O}(0)$.
This equation describes how a ``simple" operator may become increasingly ``complex" as time evolves. Then one can introduce a super-operator called {\it Liouvillian}
	\begin{equation}
		\mathcal{L}:=[H, \cdot ],
	\end{equation}
	which is a linear map in the space of operators such that
	\begin{equation}\label{LO}
		\mathcal{L}\mathcal{O}=[H,\mathcal{O}].
	\end{equation}
	Obviously we have
	\begin{equation}
		\mathcal{L}^{n}\mathcal{O}=\tilde{\mathcal{O}}_{n},
	\end{equation}
	and the Eq.\eqref{1.22} can be reformulated as
	\begin{equation}\label{1.8}
		\mathcal{O}(t)=\sum_{n=0}^{\infty}\frac{(it)^{n}}{n!}\mathcal{L}^{n}\mathcal{O}=e^{it\mathcal{L}}\mathcal{O}.
	\end{equation}
 Compared to the Schr\"odinger's picture of ordinary wave function, the above equation can be explained as the operator's ``wave function" expanded in some local basis of ``states" $\tilde{\mathcal{O}}_{n}$ belonging to a local Hilbert space $\mathcal{H}_{\mathcal{O}}$, known as {\it Krylov space}. Formally, Eq.\eqref{1.8} resembles the wave function in the Schr\"odinger picture, with $\mathcal{L}$ playing a role as the Hamiltonian. In the following, we will use the symbol $\left |A\right )$ to represent a state in the Krylov space, i.e. $|A)\in\mathcal{H}_{\mathcal{O}}$.

Since we are interested in the effects of finite temperature, we can define the inner product in Krylov space as the Wightman inner product 
	\begin{equation}\label{wightman}
		\left (A|B\right ):=\expval{e^{\beta H/2}A^{\dagger}e^{-\beta H/2}B}_{\beta}\equiv\frac{1}{\mathcal{Z}_{\beta}}\tr(e^{-\beta H/2}A^{\dagger}e^{-\beta H/2}B),
	\end{equation}
	where $\beta $ is the inverse of the temperature $T=\beta^{-1}$ and $\mathcal{Z}_{\beta}=\tr(e^{-\beta H} )$ is the thermal partition function. With the definition of the Wightman inner product, the Liouvillian $\mathcal{L}$ is Hermitian, i.e, $(A|\mathcal{L}B)=(\mathcal{L}A|B)$. Note that Eq.\eqref{1.8} can be considered as the expansion of $\left |\mathcal{O}(t)\right )$ in the basis $\{\mathcal{L}^{n}\left |\mathcal{O}\right)\} $, which are not necessarily orthonormal. However, we can use the Gram-Schmidt orthogonalization procedure \cite{gram1883ueber,schmidt1907theorie} to obtain a set of orthonormal basis $\{\left |\mathcal{O}_{n}\right )\}$ called the {\it Krylov basis} \cite{Parker:2018yvk}.  Suppose the initially given operator $\mathcal{O}$ itself is a normalized state in the Krylov space. We can define
	\begin{equation}
		\left |\mathcal{O}_{0}\right ):=\left |\mathcal{O}\right ).
	\end{equation}
	Afterwards, we can continuously construct vectors that are orthonormal to the existing basis vectors to obtain the Krylov basis. {\footnote{Besides, we assume $\mathcal{O}$ is a Hermitian operator $\mathcal{O}^\dagger=\mathcal{O}$, so that it is an observable.}} Thus, $\left |\mathcal{O}_{1}\right )$ can be constructed as follows \footnote{From the inner product Eq.\eqref{wightman} and Eq.\eqref{LO},  it is readily to get that $\left(\mathcal{O}_{0}\right |\mathcal{L}\left |\mathcal{O}_{0}\right ) =0$.}
	\begin{equation}
		b_{1}\left |\mathcal{O}_{1}\right )=\mathcal{L}\left |\mathcal{O}_{0}\right)-\left |\mathcal{O}_{0}\right )\left (\mathcal{O}_{0}\right |\mathcal{L}\left |\mathcal{O}_{0}\right ) =\mathcal{L}\left |\mathcal{O}_{0}\right ),
	\end{equation}
	where $b_{1}:=(\mathcal{L}{\mathcal{O}_0}|\mathcal{L}{\mathcal{O}_0} )^{1/2}$.  For $n>1$, $\left |\mathcal{O}_{n}\right )$ can be constructed as
	\begin{equation}\label{bn}
		b_{n}\left |\mathcal{O}_{n}\right ):=\left |A_{n}\right )=\mathcal{L}\left |\mathcal{O}_{n-1}\right )-b_{n-1}\left |\mathcal{O}_{n-2}\right ),
	\end{equation}
	where $b_{n}=(A_{n}|A_{n})^{1/2}$. The sequence $\{b_{n}\}$ is called the {\it Lanczos coefficients}. The above procedure is also known as {\it Lanczos algorithm}. 
	
	\subsection{Krylov complexity}
	The information about the growth of the operator $\mathcal{O}(t)$ is contained in the sequence Eq.\eqref{bn}. We can expand $\left |\mathcal{O}(t)\right )$ in terms of Krylov basis $\{\left |\mathcal{O}_{n}\right )\}$ as
	\begin{equation}\label{1.13}
		\left |\mathcal{O}(t)\right )=\sum_{n=0}^\infty\left |\mathcal{O}_{n}\right )\left (\mathcal{O}_{n}\right |\left .\mathcal{O}(t)\right )\equiv\sum_{n=0}^\infty i^{n}\varphi_{n}(t)\left |\mathcal{O}_{n}\right ),
	\end{equation}
	 where  $\varphi_{n}(t)$ are the {\it probability amplitudes} and satisfy
	\begin{equation}
		\sum_{n=0}^{\infty}\abs{\varphi_{n}(t)}^{2}=1.
	\end{equation}
	Combined with Eq.\eqref{1.1}, we can obtain a discrete ``Schr\"odinger" equation
	\begin{equation}\label{1.15}
		\partial_{t}\varphi_{n}(t)=b_{n}\varphi_{n-1}(t)-b_{n+1}\varphi_{n+1}(t).
	\end{equation}
	According to Eq.\eqref{1.13}, one can find that $\varphi_{n}(0)=\delta_{n,0}$ and from the recursion Eq.\eqref{1.15} we can define $\varphi_{-1}(t)\equiv0$. The Krylov complexity of an operator $\mathcal{O}$ is defined as 
	\begin{equation}\label{1.16}
		K(t):=\left (\mathcal{O}(t)|n|\mathcal{O}(t)\right )=\sum_{n=0}^{\infty}n\abs{\varphi_{n}(t)}^{2}.
	\end{equation}
    For convenience, the definition of Krylov complexity we are going to use will differ slightly from Eq.\eqref{1.16}. In order to be consistent with the Krylov complexity in \cite{Dymarsky:2021bjq}, we redefine it as
    \begin{equation}
    	K(t):=1+\sum_{n=0}^{\infty}n\abs{\varphi_{n}(t)}^{2}.
    \end{equation}
    It is worth noting that Eq.\eqref{1.15} links the growth of the operator with the hopping problem on a one-dimensional chain, where the Lanczos coefficients $b_{n}$ can be considered as the hopping amplitudes \cite{Parker:2018yvk}. Therefore, the definition of the Krylov complexity indicates that $K(t)$ represents the average position of the wave function on the chain. Hence, the growth of the operator implies an increase in the number of $n$ that contributes to $K(t)$. 
    
    In the study of Krylov complexity, there exists a very important quantity, i.e., the {\it autocorrelation function} or {\it thermal Wightman 2-point function},  
	\begin{equation}
		\begin{aligned}\label{Ct}
			C(t) & :=\varphi_{0}(t) \equiv(\mathcal{O}(t) \mid \mathcal{O}(0))  \\
			& \equiv\left\langle e^{i(t-i \beta / 2) H} \mathcal{O}^{\dagger}(0) e^{-i(t-i \beta / 2) H} \mathcal{O}(0)\right\rangle_{\beta} \\
			& =\left\langle\mathcal{O}^{\dagger}(t-i \beta / 2) \mathcal{O}(0)\right\rangle_{\beta}:=\Pi^{W}(t).
		\end{aligned}
	\end{equation}
	It is not difficult to find that as long as the derivatives of the autocorrelation function at all orders are known, the Krylov complexity can also be determined. To this end, we can expand the autocorrelation function around $t=0$,
	\begin{equation}\label{Pi}
		\begin{aligned}
			\Pi^{W}(t)&=\sum_{n=0}^{\infty}\left (\mathcal{O}\right |\frac{(-it)^{n}}{n!}\mathcal{L}^{n}\left |\mathcal{O}\right )\\
			&=\sum_{n=0}^{\infty}\mu_{2n}\frac{(it)^{2n}}{(2n)!},
		\end{aligned}
	\end{equation}
	in which we have used the facts that $(\mathcal{O}|\mathcal{L}^p|\mathcal{O})=0$ with $p$ an odd number. Therefore, only terms with even powers of $\mathcal{L}$ are left. In the above Eq.\eqref{Pi},  $\{\mu_{2n}\}$ is called the {\it moments} with the expression
 	\begin{equation}\label{mu2n}
		\mu_{2n}:=\left (\mathcal{O}\right |\mathcal{L}^{2n}\left |\mathcal{O}\right )=\left .\frac{1}{i^{2n}}\frac{d^{2n}}{dt^{2n}}\Pi^{W}(t)\right |_{t=0}.
	\end{equation}
	These moments can also be obtained from the {\it Wightman power spectrum $f^{W}(\omega)$}, which is related to the Wightman 2-point function via a Fourier transformation
	\begin{equation}\label{fW}
		f^{W}(\omega)=\int_{-\infty}^{\infty}dt e^{i\omega t}\Pi^{W}( t).
	\end{equation}
	From Eqs.\eqref{mu2n} and \eqref{fW}, the moments are related to $f^{W}(\omega)$ via 
	\begin{equation}\label{1.222}
		\mu_{2n}=\frac{1}{2\pi}\int_{-\infty}^{\infty}d\omega \omega^{2n}f^{W}(\omega).
	\end{equation}
	In particular, there is a nonlinear recursive relationship between moments and Lanczos coefficients, $b_1^{2n}\cdots b_n^2=\det\left(\mu_{i+j}\right)_{0\leq i,j\leq n}$, where $\mu_{i+j}$ is a Hankel matrix built from the moments \cite{viswanath1994recursion}. Alternatively, this expression can be reformulated in the following recursive relations, 
	\begin{gather}
		b_{n}=\sqrt{M_{2n}^{(n)}},\label{1.23}
	\end{gather}
	where,
	\begin{gather}\label{1.24}
		M^{(j)}_{2l}=\frac{M^{(j-1)}_{2l}}{b_{j-1}^{2}}-\frac{M^{(j-2)}_{2l-2}}{b^{2}_{j-2}},\qquad l=j,\dots,n,
    \end{gather}
 with $M^{(0)}_{2l}=\mu_{2l}, b_{-1}\equiv b_{0}:=1, M^{(-1)}_{2l}=0$.

	\subsection{Krylov complexity in scalar field theory with low temperatures $\beta m\gg1$}\label{sec3}
	
In this subsection, we will give a brief review on the Krylov complexity in low-temperature scalar field theory based on \cite{Camargo:2022rnt}.	Consider a real massive  scalar field $\phi$ in five-dimensional spacetime, we can write its Lagrangian in the following form
	\begin{equation}
		L_\phi=\frac{1}{2}(\partial_{\mu}\phi\partial^{\mu}\phi-m^{2}\phi^{2}),
	\end{equation} 
	where $\mu=0,1,2,3,4$ and $x^{0}=-i\tau$ with $\tau$ the Euclidean time. Then the Wightman power spectrum $f^{W}(\omega)$ can be obtained at the finite temperature $\beta^{-1}=T$ as,
	\begin{equation}\label{1.2}
		f^{W}(\omega)=\tilde N(m,\beta)(\omega^{2}-m^{2})\Theta(\abs{\omega}-m)\bigg/\sinh\left (\frac{\beta|\omega|}{2}\right ),
	\end{equation}
	where the symbol $\Theta$ represents the Heaviside step function and $\tilde N(m,\beta)$ is the normalization factor satisfying
	\begin{equation}
		\frac{1}{2\pi}\int_{-\infty}^{\infty}d\omega{f^{W}(\omega)}=1.\label{1.3}
	\end{equation}
	
	 In order to get the Krylov complexity, we need to calculate the Lanczos coefficients $b_{n}$ by solving the discrete Schr\"odinger equation \eqref{1.15}. In general, this problem is not easy to do. However, it is relatively simple at low temperatures \cite{Camargo:2022rnt}. It is worth noting that in finite temperature field theories, low temperature means $\beta m\gg 1$, i.e., temperature is much less than the mass of the field, rather than simply $\beta\gg 1$ \cite{Kapusta:2006pm}. Therefore, even if $\beta$ is not very large, it may still be considered as low temperature if $m$ is sufficiently large. 
	In the low temperature limit $\beta m\gg 1$,  we can rewrite the Wightman power spectrum Eq.\eqref{1.2} as
	\begin{equation}\label{1.5}
		f^{W}(\omega)\approx N(m,\beta)e^{-\beta\abs{\omega}/2}(\omega^{2}-m^{2})\Theta(\abs{\omega}-m), \qquad \beta m\gg1,
	\end{equation}
	in which we have absorbed the extra term $2$ into the normalization factor $N(m,\beta)$, i.e., $N(m,\beta)=2\tilde N(m,\beta)$. Using Eq.\eqref{1.3}, $N(m,\beta)$ can be determined analytically,
	\begin{equation}
		N(m,\beta)=\frac{\pi\beta^{3}e^{\beta m/2}}{16+8\beta m}, \qquad  ~\beta m\gg1.
	\end{equation}
	Therefore, the moments $\mu_{2n}$ in Eq.\eqref{1.222} can be directly computed as
	\begin{equation}\label{1.7}
		\mu_{2n}=\frac{2^{-2}e^{\frac{\beta m}{2}}}{2+\beta m}\left (\frac{2}{\beta}\right )^{2n}\left [4\tilde{\Gamma}\left (3+2n,\frac{\beta m}{2}\right )-\beta^{2}m^{2}\tilde{\Gamma}\left (1+2n,\frac{\beta m}{2}\right )\right ],
	\end{equation}
	where $\tilde{\Gamma}(n,z)$ is the incomplete Gamma function. 
	
\begin{figure}[htb]
	\centering
	\includegraphics[width=0.6\linewidth]{"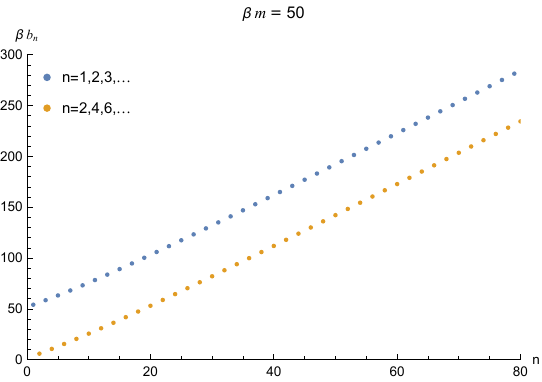"}
	\caption{Lanczos coefficients $b_{n}$ for { $\beta m=50$}. The blue dots are for odd $n$ while the brown dots are for even $n$.}
	\label{fig:low-bn}
\end{figure}

	Following the non-linear recursive relation in Eqs.\eqref{1.23}-\eqref{1.24}, the Lanczos coefficients $\{b_{n}\}$ can be obtained from the Eq.\eqref{1.7}. For example, we show the behavior of the Lanczos coefficients in Figure \ref{fig:low-bn} with { $\beta m=50$}. We see that $b_n$ exhibits staggering behavior, meaning that it is divided into two groups: one is with odd $n$ (blue dots) while the other is with even $n$ (brown dots). Each group is a monotonically increasing function with respect to $n$. In \cite{Camargo:2022rnt} the authors also observed that the separation $\Delta b_n:=|b_n^{\rm odd}-b_n^{\rm even}|$ is of the order of the mass $m$.
	
The Krylov complexity can be computed from the auto-correlation function $\varphi_{0}(t)$ in Eq.\eqref{Ct} and the Lanczos coefficients by means of Eq.\eqref{1.15}. Performing the Fourier transformation of $f^{W}(\omega)$ one can yield the auto-correlation function as
\begin{equation}
\begin{aligned}\label{varphi0}
		\varphi_{0}(t)&=\frac{1}{2\pi}\int_{-\infty}^{\infty}d\omega e^{-i\omega t}f^{W}(\omega)\\
		&=\frac{1}{\pi}\int_{0}^{\infty}d\omega \cos(\omega t) f^{W}(\omega)\\
		&=\frac{\beta^{3}\left((\beta^{3}(2+\beta m)-24\beta t^{2}-16mt^{4})\cos(mt)-4t(\beta^{2}(3+\beta m)+4(\beta m-1)t^{2})\sin(mt)\right)}{(2+\beta m)(\beta^{2}+4t^{2})^{3}}.
\end{aligned}
\end{equation}
in which we have used the fact that $\varphi_0(t)$ is a real function and $f^W(\omega)$ is an even function of $\omega$. Afterwards, we can use the fourth Runge-Kutta method to numerically solve Eq.\eqref{1.15} and consequently obtain the Krylov complexity $K(t)$ and the Krylov entropy $S_K(t)$, which is defined by \cite{Barbon:2019wsy}
\begin{equation}
	S_{K}(t):=-\sum_{n=0}^{\infty}\abs{\varphi_{n}(t)}^{2}\log\abs{\varphi_{n}(t)}^{2}.
\end{equation}
\begin{figure}[htb]
	\centering
	\includegraphics[width=0.45\linewidth]{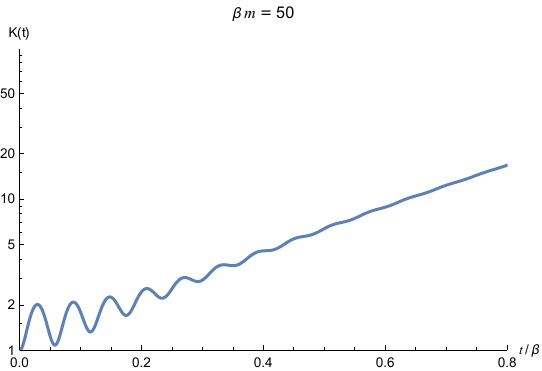}~~
	\includegraphics[width=0.45\linewidth]{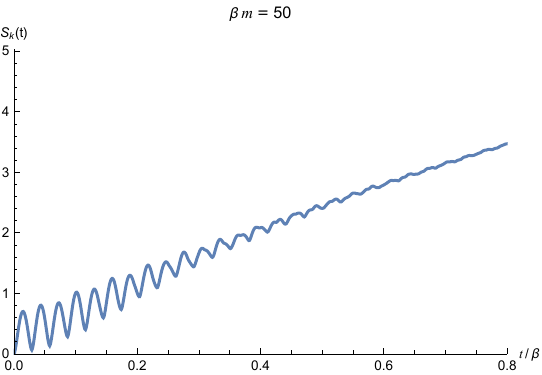}
	\caption{Time evolution of Krylov complexity (left panel) and Krylov entropy (right panel) of the free scalar field theories for { $\beta m=50$} in five dimensional spacetime. The vertical axis is in a logarithmic scale for $K(t)$ in the left panel. }\label{fig:low-krylov}
\end{figure}

We show the time evolution of $K(t)$ and $S_K(t)$ in Figure \ref{fig:low-krylov} as an example with { $\beta m=50$}. In the left panel of Figure \ref{fig:low-krylov}, it is found that the Krylov complexity oscillates when time is short. As was explained in \cite{Camargo:2022rnt}, these oscillations are due to the trigonometric functions in the auto-correlation function \eqref{varphi0}. The oscillatory behavior of $\varphi_0(t)$ will subsequently pass on to $\varphi_n(t)$ according to the discrete Schr\"odinger equation \eqref{1.15}. In the early time the period of the oscillations is roughly $\pi/m$, which is consistent with the plots in the left panel of Figure \ref{fig:low-krylov} where the period is roughly $0.06$.  In the late time, the amplitudes of the oscillations becomes smaller, which is due to the cancellation of $\varphi_n(t)$ coming from various $n$. However, in the early time only small number of $n$ will contribute to $\varphi_n(t)$. Thus, the oscillation in the early time is more significant than in the late time.  Intuitively, it might seem difficult to comprehend this oscillatory behavior, as a `simple' operator will evolve towards a more `complex' one \cite{Parker:2018yvk}. However, in fact, the operator becoming more `complex' is directly related to $K(t)$ which receives contributions from more number of $n$. Therefore, $K(t)$ as the position of the wave function on the one-dimensional chain, is not necessarily monotonically increasing. In the following subsection \ref{sec4.1} we will explain in detail whether the position of the wave function increases monotonically or not is related to the behavior of the hopping amplitude $b_n$ as well as the wave function $\varphi_0(t)$. 

The above calculations only applies to the low temperature case of $\beta m\gg1$ since Eq.\eqref{1.5} merely works for $\beta m\gg1$. Therefore, we cannot use it for more general temperatures. In the next section, we will provide detailed studies of Krylov complexity and Krylov entropy for more general situations of $\beta m$.

\section{Krylov complexity and Krylov entropy with General Temperatures}\label{sec4}
In this section, we will relax the constraint on $\beta m$ to study the Krylov complexity and Krylov entropy in a more general situation. The key idea is that we can rewrite the Wightman power spectrum $f^W(\omega)$ \eqref{1.2} in a series of summation with general $\beta m$, which subsequently makes the computation of Krylov complexity and Krylov entropy feasible. 

\subsection{Truncations of the Wightman power spectrum}
For general values of $\beta m$, the Wightman power spectrum \eqref{1.2} can be rewritten as an infinite series. For convenience, we consider only the part for $\omega>m$ and the other part for $-\omega>m$ can be obtained from the fact that Eq.\eqref{1.2} is an even function with respect to $\omega$. Therefore, for $\omega>m$ we have
\begin{equation}
	\begin{aligned}
		f^{W}(\omega)&=\tilde N(\beta,m)(\omega^{2}-m^{2})/\sinh(\frac{\beta\omega}{2})\\
		&=2\tilde N(\beta,m)(\omega^{2}-m^{2})\frac{e^{-\frac{\beta\omega}{2}}}{1-e^{-\beta\omega}}\\
		&=2\tilde N(\beta,m)(\omega^{2}-m^{2})e^{-\frac{\beta\omega}{2}}\sum_{k=0}^\infty e^{-\beta\omega k}\\
		&=2\sum_{k=0}^{\infty}\tilde N(\beta,m)(\omega^{2}-m^{2})e^{-\beta\omega(k+1/2)},
	\end{aligned}
\end{equation}
in which we have used the series expansion $\frac{1}{1-x}=\sum_{k=0}^\infty x^k$. Let $\beta_{k}\equiv\beta(2k+1)$ and absorb the factor $2$ into $N(\beta,m)$ we obtain
\begin{equation}\label{3.2}
	f^{W}(\omega)=N(\beta,m)\sum_{k=0}^{\infty}(\omega^{2}-m^{2})e^{-\beta_{k}\omega/2},
\end{equation}
 where $N(\beta,m)$ can be calculated from the normalization conditions \eqref{1.3} and its exact form is
\begin{equation}\label{2.3}
	N(\beta,m)=\frac{\pi\beta^{3}e^{\frac{3\beta m}{2}}}{2\beta m\Phi(e^{-\beta m},2,3/2)+2e^{\beta m}(4\beta m+\Phi(e^{-\beta m},3,1/2))},
\end{equation}
in which $\Phi(z,s,a)\equiv\sum_{k=0}^{\infty}z^{k}/(k+a)^{s}$ is the Lerch transcendent function \cite{lerch}. Therefore, the series of the Wightman power spectrum in Eq.\eqref{3.2} can in principle work for general $\beta m$'s. It has much broader applications than that in Eq.\eqref{1.5} from \cite{Camargo:2022rnt}, which can only apply in the low temperature limit.

\begin{figure}
	\centering
	\includegraphics[width=0.45\linewidth]{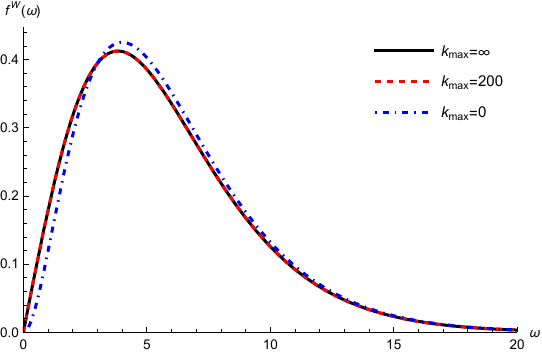}~~
	\includegraphics[width=0.45\linewidth]{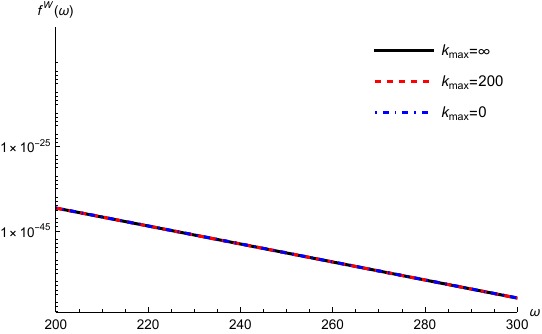}
	\caption{Relation between the Wightman power spectrum $f^{W}(\omega)$ against the frequency $\omega$ with $\beta=1$ and $m=0.01$. (Left) The black line ($k_{\text{max}}=\infty$) corresponds to Eq.\eqref{3.2}, while the blue dash-dotted line ($k_{\text{max}}=0$) corresponds to Eq.\eqref{1.5}. The two lines deviate from each other as $\omega\lesssim 12$. The red dashed line ($k_{\text{max}}=200$) corresponds to Eq.\eqref{3.10}. It matches with the black line in the whole regime of the frequency, reflecting that taking $k_{\text{max}}=200$ is accurate enough for the numerical calculations; (Right) The logarithmic plot of $f^W(\omega)$ against $\omega$ in the limit of large frequency. The three lines (black, red dashed and blue dash-dotted) for different $k_{\rm max}$ collapse together and the green dashed line is the best fit of them. They share the same scaling law as $f^W(\omega)\sim 1611.14\times e^{-0.4919\omega}$.  }\label{fig:f2}
\end{figure}

 In order to compute the moments $\mu_{2n}$ \eqref{1.222} we first define
\begin{equation}\label{3.5}
	I(n)\equiv \int_{m}^{\infty}\omega^{n}e^{-\frac{\beta_{k}\omega}{2}}d\omega
\end{equation}
which can be integrated by parts as, 
\begin{gather}
	I(n)=\frac{2}{\beta_{k}}e^{-\frac{\beta_{k}m}{2}}m^{n}+\frac{2n}{\beta_{k}}I(n-1).
\end{gather}
In particular, for $n=0$ we can get it directly from the integration Eq.\eqref{3.5},
\begin{gather}
	I(0)=\int_{m}^{\infty}e^{-\frac{\beta_{k}\omega}{2}}d\omega=\frac{2}{\beta_{k}}e^{-\frac{\beta_{k}m}{2}}.
\end{gather}
These equations imply that
\begin{equation}
	 \begin{aligned}
		 I(n)&=\left (\frac{2}{\beta_{k}}\right )^{n+1}\tilde{\Gamma}\left (n+1,\frac{\beta_{k}m}{2}\right ).
	\end{aligned}
\end{equation}
Therefore, from the definition of moments Eq.\eqref{1.222}, we have
\begin{equation}\label{sum}
	\begin{aligned}
		\mu_{2n}&=\int_{-\infty}^{\infty}\frac{\omega^{2n}}{2\pi}\left [N(\beta m)\sum_{k=0}^{\infty}(\omega^{2}-m^{2})e^{-\beta_{k}\omega/2}\right ]\\
		&=\frac{N(\beta,m)}{\pi}\sum_{k=0}^{\infty}\left [I(2(n+1))-m^{2}I(2n)\right ]\\
		&=N(\beta,m)\sum_{k=0}^{\infty}\frac{2^{2n+1}\beta_{k}^{-(2n+3)}}{\pi}\left [4\tilde{\Gamma}\left (2n+3,\frac{\beta_{k}m}{2}\right )-\beta_{k}^{2}m^{2}\tilde{\Gamma}\left (2n+1,\frac{\beta_{k}m}{2}\right )\right ].
	\end{aligned}
\end{equation}
Unfortunately, this formula is too complex to be computed analytically. Therefore, we resort to numerical computations.
To this end, we can truncate the summation of $\mu_{2n}$ in Eq.\eqref{sum} at a suitable $k$ to obtain approximate results. In order to ensure sufficient computational speed and accuracy, the truncation position of $k$ should not be too large or too small. We find that a suitable choice is to make the normalization factor of the truncated Wightman power spectrum close to the exact form in Eq.\eqref{2.3}. In the remainder of this paper, we always take
\begin{equation}\label{3.10}
	f^{W}\approx N(\beta,m)\sum_{k=0}^{k_{\text{max}}}(\omega^{2}-m^{2})e^{-\beta_{k}\omega/2},
\end{equation}
in which $k_{\rm max}$ is a large cut-off of $k$. We can see that the exact $f^{W}(\omega)$ in Eq.\eqref{3.2} satisfies $k_{\text{max}}=\infty$, while the approximated $f^{W}(\omega)$ in Eq.\eqref{1.5} in the low temperature limit satisfies $k_{\text{max}}=0$. In our case we choose $k_{\text{max}}=200$ in Eq.\eqref{3.10}, because in most cases it is sufficient to obtain accurate numerical results. In Figure \ref{fig:f2} we compare the exact Wightman power spectrum $f^W$ to the approximated functions with $\beta=1$ and $m=0.01$, i.e. $\beta m\ll1$ in the limit of high temperature. In the left panel of Figure \ref{fig:f2}, we see that $f^{W}(\omega)$ for $k_{\rm max}=0$ (blue dash-dotted line) is only consistent with the exact form (black line with $k_{\rm max}=\infty$) in the regime of large frequency; However, they will deviate from each other as $\omega\lesssim 12$, which means the formula in Eq.\eqref{1.5} is not accurate in the low frequency regime with high temperatures. On the contrary, still in the left panel of Figure \ref{fig:f2}, we can see that there is no distinguishable differences for the Wightman power spectrum $f^{W}(\omega)$ between $k_{\text{max}}=\infty$ (black line) and $k_{\text{max}}=200$ (red dashed line) in the whole regime of the frequency, which verifies that our choice of $k_{\text{max}}=200$ is sufficient for the accuracy.

In the right panel of Figure \ref{fig:f2} we show the logarithmic relations between $f^W(\omega)$ and $\omega$ in the large frequency limit.  We can find that in the large frequency limit, the Wightman power spectrum with various $k_{\rm max}$'s will overlap together and decay exponentially with the frequency, 
\be\label{fitfW}
f^W(\omega)\sim e^{-\omega/\omega_0}, \qquad ~~\omega\to\infty.
\ee
The fitted line (green dashed line) in the right panel of Figure \ref{fig:f2} shows that this decay rate is $1/\omega_0\approx 0.4919$ as $\omega\to\infty$. Therefore, we get $\omega_0\approx 2.0329$, which is consistent with discussions in \cite{Camargo:2022rnt} that the leading exponential decay of the power spectrum is $\omega_0=2/\beta$ where we have set $\beta=1$. 


\subsection{Lanczos coefficients with truncations}
Before studying the Krylov complexity, let's first look at the behaviors of the Lanczos coefficients $b_{n}$ with the truncations $k_{\rm max}=200$. 

\begin{figure}[htbp]
	\centering
	\includegraphics[width=0.6\linewidth]{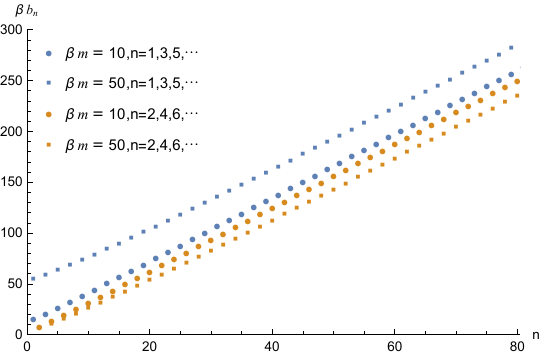}
	\caption{Lanczos coefficients $b_{n}$ for { $\beta m=10$} (dots) and { $\beta m=50$} (squares) with $k_{\text{max}}=200$. They are grouped in two families with odd $n$ (blue) and even $n$ (brown).}
	\label{fig:bn10}
\end{figure}

In Figure \ref{fig:bn10}, we present the behaviors of $b_{n}$ for { $\beta m=10$} (dots) and { $\beta m=50$} (squares).
 These coefficients also exhibit clear behavior of ``staggering", which are grouped in two families with odd (in blue) and even $n$ (in brown). As $n$ is relatively large, $b_n$ becomes linearly proportional to $n$. Intuitively, we can find that the separation $\Delta b_n$ between $b_n$ for the odd $n$ and even $n$ with { $\beta m=50$} are greater than those with { $\beta m=10$}. This phenomenon is consistent with the discussions in the preceding section that $\Delta b_n$ is of the order of the mass $m$ \cite{Camargo:2022rnt}. In order to analyze them quantitatively, we fit $b_{n}$ linearly for odd $n$ and even $n$ separately similar to \cite{Camargo:2022rnt}:
\begin{gather}
	b_{n}\sim \alpha_{\text{odd}}n+\gamma_{\text{odd}},\qquad \text{odd $n$},\label{3.11}\\
	b_{n}\sim \alpha_{\text{even}}n+\gamma_{\text{even}},\qquad \text{even $n$}\label{3.12},
\end{gather}
where $\alpha_{\text{odd}},\alpha_{\text{even}},\gamma_{\text{odd}}$ and $\gamma_{\text{even}}$ are constants which are independent of $n$. Specifically, we perform the linear fitting for $b_n$ in the range of $n\in [100,450]$. 

\begin{figure}[htb]
	\centering
	{\includegraphics[width=0.45\linewidth]{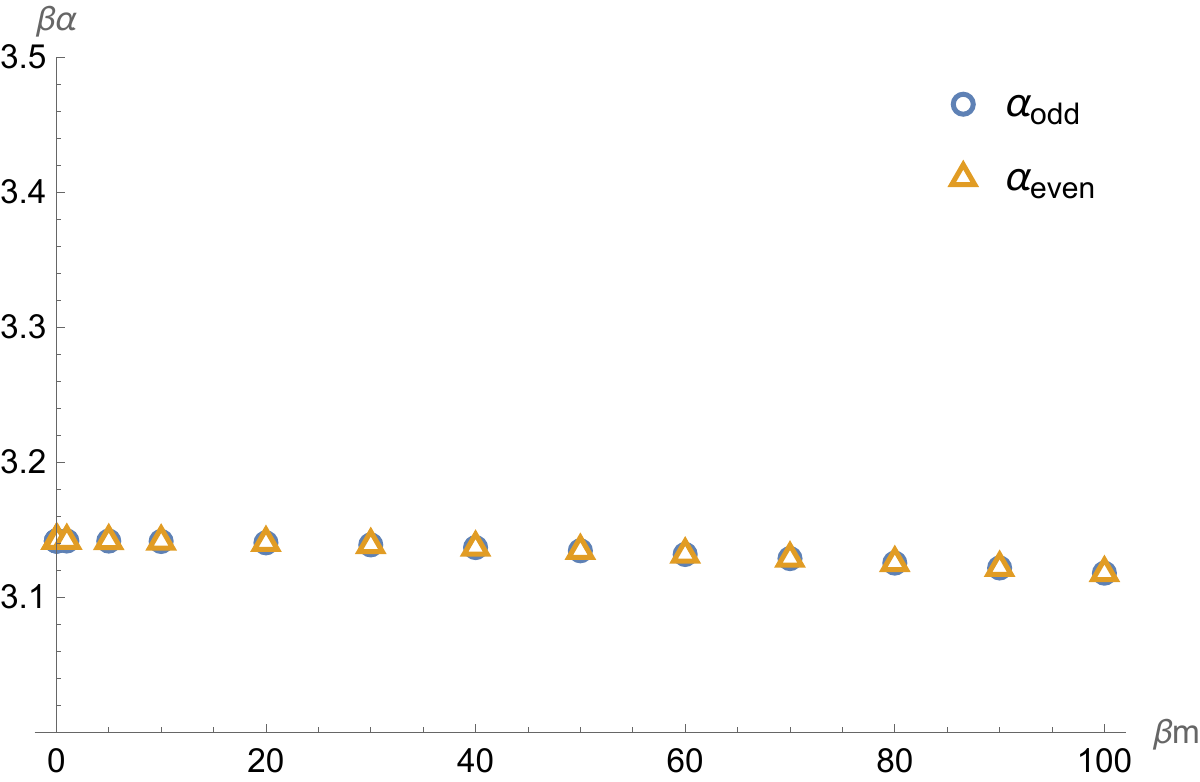}}~~
	{\includegraphics[width=0.45\linewidth]{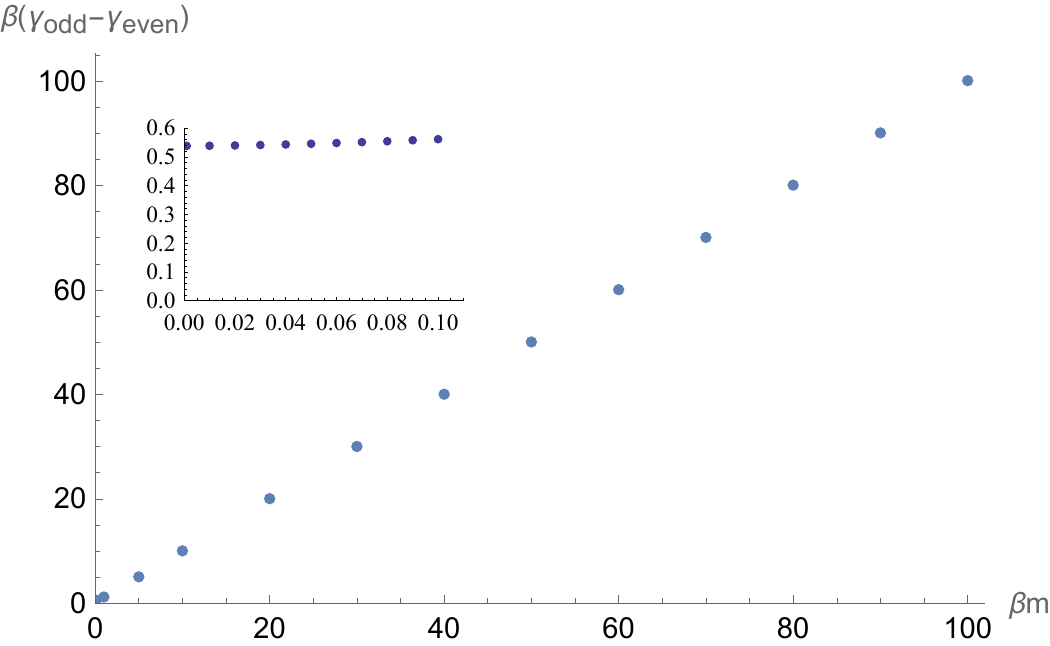}}
	\caption{(Left) Mass-dependence of the slope $\alpha_{\text{odd}}$ and $\alpha_{\text{even}}$ in Eqs.\eqref{3.11}-\eqref{3.12}; (Right) Mass-dependence of the difference $\gamma_{\text{odd}}-\gamma_{\text{even}}$ in Eqs.\eqref{3.11}-\eqref{3.12}. The inset plot exhibits the difference in high-temperatures regime. }
	\label{fig:slope}
\end{figure}

In the left panel of Figure \ref{fig:slope}, we show the relation between $\beta \alpha$ ($\alpha$ means $\alpha_{\rm odd} $ (blue) and $\alpha_{\rm even}$ (brown)) against $\beta m$. It is not difficult to observe that the two coefficients $\alpha_{\text{odd}}$ and $\alpha_{\text{even}}$ will overlap exactly, meaning that the slope in the fitting relations Eq.\eqref{3.11} and Eq.\eqref{3.12} are exactly identical. Moreover, we see that as $\beta m\to0$ the value of $\alpha_{\text{odd}}$ and $\alpha_{\text{even}}$ are approaching $\pi/\beta$. However, as $\beta m$ increases, $\alpha_{\text{odd}}$ and $\alpha_{\text{even}}$ slightly decrease from $\pi/\beta$. This may come from the selected range of $n$ when fitting $b_n$ as argued in \cite{Camargo:2022rnt}. It is expected that larger range of $n$ will improve the value of $\alpha$, and it will finally approach $\pi/\beta$ more closely \cite{Camargo:2022rnt}. Therefore, in our case we can speculate that $\alpha_{\text{odd}}=\alpha_{\text{even}}\approx\pi/\beta$. This analysis is also consistent with the {\it universal operator growth hypothesis}  in \cite{Parker:2018yvk} that the Lanczos coefficients $b_n$ in a generic chaotic system will grow as fast as $b_n\sim\alpha n+\gamma$ with $\alpha=\pi\omega_0/2$ as $n\to\infty$. In the fitting of the Wightman power spectrum Eq.\eqref{fitfW} we already get $\omega_0\approx2/\beta$, therefore, we can get $\alpha\approx\pi/\beta$ which is consistent with the left panel of Figure \ref{fig:slope}.

In the right panel of Figure \ref{fig:slope}, we show the relation between the difference $\beta(\gamma_{\text{odd}}-\gamma_{\text{even}})$ with respect to $\beta m$.  For larger $m$, we can see that $\beta(\gamma_{\text{odd}}-\gamma_{\text{even}})$ is linearly proportional to $\beta m$ with the ratio $\beta(\gamma_{\text{odd}}-\gamma_{\text{even}})/(\beta m)\approx 1$, which is consistent with the separations of $b_n$ in Figure \ref{fig:bn10} that $\Delta b_n$ is of the order of the mass $m$.  However, we find that this proportionality will break down for small $\beta m\in[0, 0.1]$, as shown in the inset plot in the right panel of Figure \ref{fig:slope}. In the inset plot, the ratio in the range of small $\beta m$ is roughly $\beta(\gamma_{\text{odd}}-\gamma_{\text{even}})/(\beta m)\approx 0.2$ which is much smaller than the ratio $1$ in the large $\beta m$ regime.  In the region of small $\beta m\ll1$, the system is in the high temperature limit, therefore, we speculate that the linear proportionality between $\gamma_{\text{odd}}-\gamma_{\text{even}}$ and $m$ may not apply due to the high temperatures. As far as we know, this phenomenon is not explored previously.

Therefore, we see that high temperature limit will bring out different phenomena from the studies in \cite{Camargo:2022rnt} with low temperatures. Next, we will first study the Krylov complexity and Krylov entropy in high-temperature limit, i.e., $\beta m\ll 1$ and then discuss the general cases with general $\beta m$. After that, we apply the results to the finite-temperature Higgs scalar field theory. 

\subsection{High-temperature limit with $\beta m\ll 1$}\label{sec4.1}

For convenience, in the numerics we always keep $\beta\equiv1$ and study the variations of the Krylov complexity and Krylov entropy with respect to $\beta m$ by adjusting the values of $m$. The high temperature limit implies $\beta m\ll 1$ \cite{Kapusta:2006pm}, therefore, the approximation of Wightman power spectrum in Eq.\eqref{1.5} from \cite{Camargo:2022rnt} is in invalid in this case. We need to use Eq.\eqref{3.10} to numerically study the Wightman power spectrum and then to study other quantities such as Krylov complexity and Krylov entropy. 

\begin{figure}[htb]
	\centering
	{\includegraphics[width=0.45\linewidth]{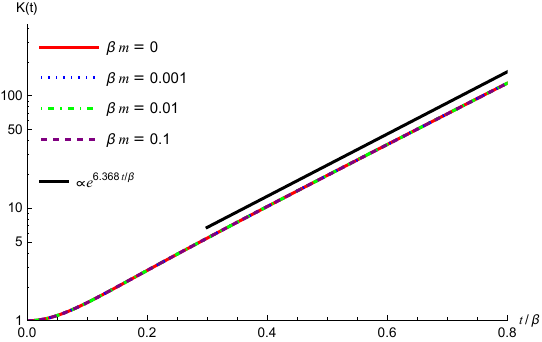}}~~
	{\includegraphics[width=0.45\linewidth]{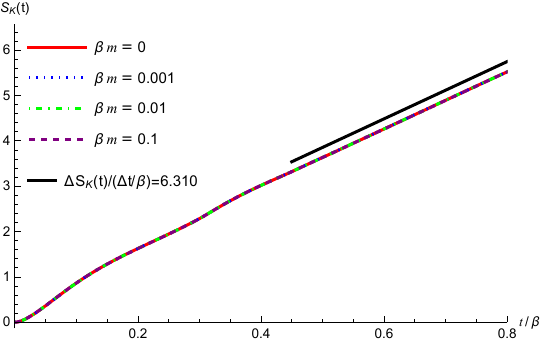}}
	\caption{(Left) Time evolution of Krylov complexity for { $\beta m\ll1$} in the logarithmic scale. They overlap with each other and scale as { $e^{6.368t/\beta}$} (as the black reference line indicates) at late time; (Right) The corresponding time evolution of Krylov entropy and the black line is the reference line with scalings { $\Delta S_K(t)/(\Delta t/\beta)=6.310$}.}
	\label{fig:hightemp}
\end{figure}

As examples, we will take the cases of { $\beta m=0, 0.001$, $0.01$} and $0.1$. We have plotted the time evolution of Krylov complexity $K(t)$ (in logarithmic scale) and Krylov entropy $S_K(t)$ in the Figure \ref{fig:hightemp}. In both panels, the red lines for { $\beta m=0$} are taken from the conformal field theory in \cite{Dymarsky:2021bjq}. From the figure, we can see that these curves (i.e., { $\beta m=0.001$}, $0.01$, $0.1$ and $0$) will overlap together and cannot be distinguished from each other. This is consistent in numerics itself since in this case { $\beta m$} are very small and close to zero.  Figure \ref{fig:hightemp} indicates that at high temperatures $\beta m\ll1$, the behaviors of the Krylov complexity and Krylov entropy for a free scalar field theory are quite similar to those of the conformal field theory.
 
From the discussions in \cite{Parker:2018yvk}, Krylov complexity will grow exponentially at late time as 
\be K(t)\propto e^{\lambda_K t}, \ee
in which $\lambda_K=\pi\omega_0=2\pi/\beta$.
In the left panel of Figure \ref{fig:hightemp}, the black line is the reference line for the Krylov complexity at late time. Therefore, we see that in the high temperature limit, the exponential growth rate for Krylov complexity is roughly $\lambda_K\approx 6.368$ which is a little bit greater than $2\pi/\beta$. This is because we are fitting the line in a finite time region. For larger time, the slope of the exponential growth will tend to $2\pi/\beta$ which is consistent with the discussions in \cite{Camargo:2022rnt,Dymarsky:2021bjq}.

In the right panel of Figure \ref{fig:hightemp} the black line is the reference line and scale as { $\Delta S_K(t)/(\Delta t/\beta)=6.310$}, indicating the linear growth of the Krylov entropy in late time. The linear slope is also close to $\lambda_K$, which is consistent with the results in \cite{Fan:2022xaa,Camargo:2022rnt} that $S_K(t)\sim\log K(t)$ at late time.

{
\subsection{Origins of the absence of oscillations in the high-temperature limit}

In the high-temperature limit, the behavior of the Krylov complexity and Krylov entropy is significantly different from that observed at low temperatures. Specifically, the oscillations that are present at low temperatures are absent at high temperatures. From the discrete Schr\"odinger equation, we can speculate that the behavior of $\varphi_n(t)$ will depend on the hopping amplitudes $b_n$ and the value of the wave function $\varphi_0(t)$. The hopping amplitudes $b_n$ describes how the wave function can `jump' from one position to the next position. Therefore, if $b_n$ is staggering between even and odd positions, we can intuitively speculate that this staggering behavior will certainly affect the monotonic increase of $\varphi_n(t)$, and further affects the behavior the Krylov complexity. Moreover, the recursive relation of $\varphi_n(t)$ in the discrete Schr\"odinger equation indicates that the behavior of $\varphi_n(t)$ will also depend on $\varphi_0(t)$. Therefore, the behavior of $\varphi_0(t)$ will also affect the behavior of $\varphi_n(t)$ as well as the Krylov complexity. In the following we will compare the behaviors of $b_n$ and $\varphi_0(t)$ in the high-temperature and in the low-temperature regions, and we will see that the numerical results are consistent with our speculations. 
	

	\begin{itemize}
		\item 
		{ Dependence on $\varphi_0(t)$:} In the Figure \ref{fig:vaphi0}, we have plotted the profiles of the $\varphi_0(t)$ for various values of $\beta m$, i.e., $\beta m=0.01$, $\beta m=10$ and $\beta m=50$. We can observe that in the low-temperature ($\beta m=50$) regime, $\varphi_0(t)$ (green line) has significant oscillations. However, on the contrary, in the high-temperature ($\beta m=0.01$) regime, $\varphi_0(t)$ (blue line) does not oscillate! Therefore, from the behavior of $\varphi_0(t)$, we can conclude that the lower the temperature is, the more oscillations of $\varphi_0(t)$ will become. Therefore, in the high-temperature regime, the absence of oscillations of Krylov complexity can be attributed to the lack of oscillations in $\varphi_{0}(t)$.
		
		
		
		\begin{figure}[h]
			\centering
			\includegraphics[width=0.6\linewidth]{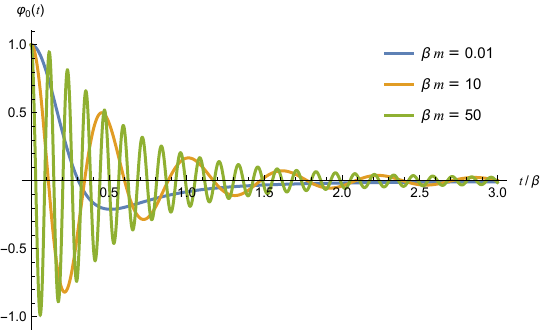}
			\caption{Behaviors of $\varphi_0(t)$ for various temperatures $\beta m$. In the low-temperature regime (green line), $\varphi_0(t)$ has significant oscillation; However, in the high-temperature regime (blue line), there are no oscillations of $\varphi_0(t)$. }\label{fig:vaphi0}
		\end{figure}

		\item {Dependence on $b_n$:}  It is known that if $b_n$ is linearly related to $n$, then the Krylov complexity grows exponentially \cite{Parker:2018yvk}. However, in our case $b_n$ separates into two groups of linear dependence on $n$, i.e., there is staggering behavior of $b_n$. Therefore, we can speculate that the staggering behavior of $b_n$ may cause the Krylov complexity to oscillate as well. 
		Let $\gamma_{1}\equiv\beta (\gamma_{\text{odd}}+\gamma_{\text{even}})$, $\gamma_{2}\equiv\beta (\gamma_{\text{odd}}-\gamma_{\text{even}})$, then we can reformulate $b_n$ in Eqs.\eqref{3.11}-\eqref{3.12} as, 
		\begin{equation}\label{eqbn}
			 \beta b_{n}=\beta \alpha n+\dfrac{\gamma_{1}}{2}+(-1)^{n+1}\frac{\gamma_{2}}{2},~~~~n\in Z.
		\end{equation}
		When $n$ is sufficiently large, the linear term becomes significant. Then, the third term on the right side of Eq.\eqref{eqbn} can be neglected. Therefore,  at this point, the Krylov complexity should grow exponentially. To clarify why there is no oscillation at high temperatures and why the oscillation disappears over time at low temperatures, we can make the function $\varphi_n(t)$ continuous  such that
			\begin{equation}
				\varphi(x,t):=\varphi_{n}(t),\qquad x=\epsilon n,
			\end{equation}
			where $\epsilon$ is a small lattice cutoff. Assuming that when $n > \hat{n}$, the third term of $\beta b_n$ in Eq.\eqref{eqbn} can be neglected, and let $\hat{x} = \epsilon \hat{n}$. Then, the Krylov complexity can also be rewritten in a continuous form
			\begin{eqnarray}\label{5}
					K(t)&=&1+\sum_{n=0}^{\infty}n\abs{\varphi_{n}(t)}^{2}\nonumber\\
					&=&1+\frac{1}{\epsilon^{2}}\int_{0}^{\infty}dx ~x\abs{\varphi(x,t)}^{2}\nonumber\\
					&=&1+\frac{1}{\epsilon^{2}}\int_{0}^{\hat{x}}dx~x\abs{\varphi(x,t)}^{2}+\frac{1}{\epsilon^{2}}\int_{\hat{x}}^{\infty}dx~x\abs{\varphi(x,t)}^{2}.
			\end{eqnarray}
			The Krylov complexity represents the position of the wave function $\varphi(x,t)$ on the Krylov chain. When $t$ is small, its position is relatively close to the origin $(x,t)=(0,0)$, and the Krylov complexity is dominated by the second term in \eqref{5}, which can disrupt the exponential growth and potentially lead to oscillations. When $t$ is sufficiently large, the position of the wave function is far from the origin, and the Krylov complexity is dominated by the third term in equation \eqref{5}, thus exhibiting exponential growth behavior. This is why oscillations gradually fade away over time in low-temperature conditions. For a detailed discussion on this topic, please refer to reference \cite{Barbon:2019wsy}.

		In the high-temperature limit, the Krylov complexity exhibits exponential growth from the very beginning. From our speculations, then the third term of $\beta b_n$ should be negligible when $n$ is very small. Therefore, in the high-temperature limit and when $n$ is very small, it should satisfy $\gamma_2/2 \ll \beta\alpha n+\gamma_{1}/2$. 
		This corresponds exactly to the high-temperature region as shown in the right panel of Figure \ref{fig:slope} and in the Figure \ref{fig:gamma1} below. In the high-temperature region, $\gamma_1$ is approximately around 6.3 (as $\beta m$ is small in the Figure \ref{fig:gamma1}), and $\gamma_2$ is less than 0.6 (see the inset plot in the right panel of Figure \ref{fig:slope}), i.e., $\gamma_2/2 \ll \beta\alpha n+\gamma_{1}/2$ for $n$ is small. 
			\begin{figure}[h]
			\centering
			\includegraphics[width=0.6\linewidth]{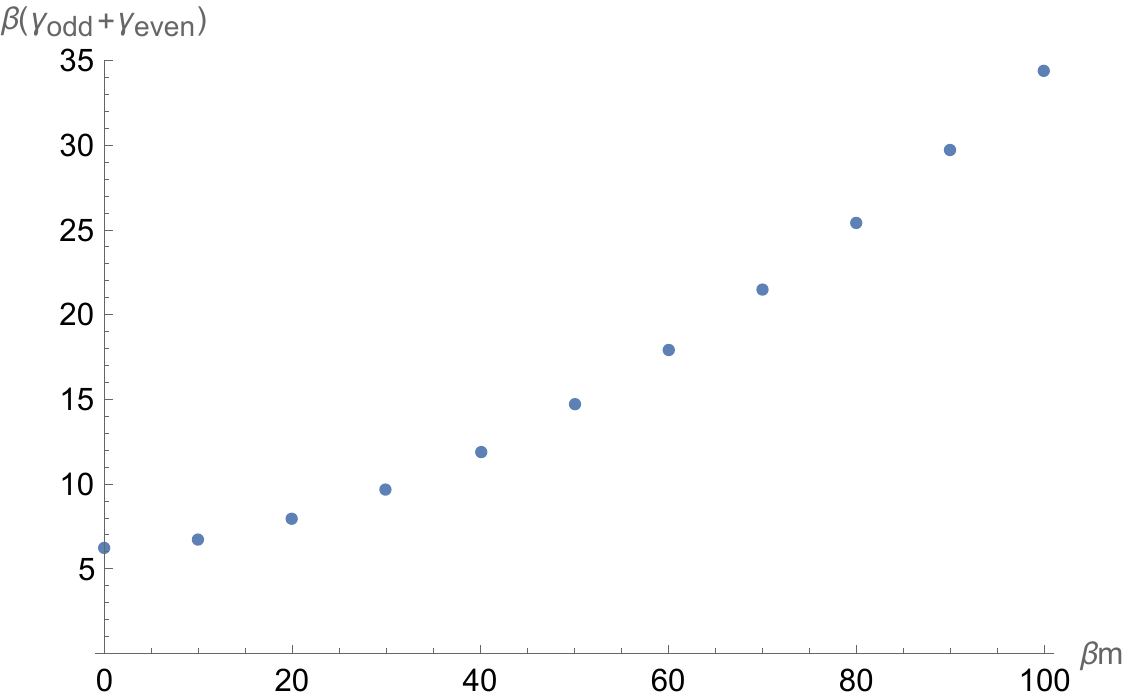}
			\caption{Relation between $\beta\gamma_1$ and the value of $\beta m$. We can see that in the high-temperature regime ($\beta m$ is small), $\beta\gamma_1\approx 6.3$. }\label{fig:gamma1}
		\end{figure}
		Therefore, in high-temperature regime, Krylov complexity does not have oscillations. However, in the low-temperature regime (as $\beta m$ is big enough) and when $n$ is small, $\gamma_2$ cannot be neglected. As we can see from the right panel of Figure \ref{fig:slope} and the Figure \ref{fig:gamma1}, when $\beta m=100$, $\beta \gamma_1$ is roughly $35$ and $\beta \gamma_2$ is roughly $100$ which cannot be neglected. This means the staggering behavior of $b_n$ cannot be neglected in low-temperature regime. Therefore, the Krylov complexity will exhibit oscillations in the low-temperature as $n$ is small.
		\end{itemize}
			
			\begin{figure}[h]
		\centering
		\includegraphics[width=0.6\linewidth]{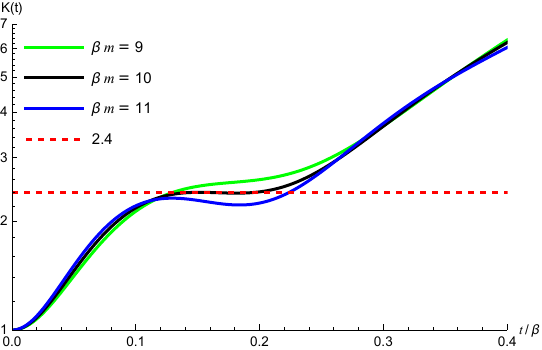}
		\caption{The Krylov complexity for $\beta m = 9, 10, 11$. As $\beta m=10$ (black line), there exists a plateau at round $K(t)\approx 2.4$ (red dashed line), which implies that $\beta m=10$ is the transition temperature that the Krylov complexity will change its behavior from monotonic increase to oscillations.}\label{fig:criticaltemperature}
	\end{figure}
		
It will be interesting to see whether theres exists a transition temperature that Krylov complexity will change its behavior from the oscillations to monotonic increase. After careful seeking, we numerically found that at around $\beta m=10$ there indeed exists such kind of transition temperature. We plot the profile of $K(t)$ as $\beta m=9,10,11$ in the Figure \ref{fig:criticaltemperature}. We observe that at this transition temperature there exists a plateau at $K(t)\approx 2.4$ (red dashed line). The existence of plateau indicates that $K(t)$ at $\beta m=10$ is a critical line, since as $\beta m<10$ the Krylov complexity will increase monotonically and while $\beta m>10$ the Krylov complexity will have oscillation behaviors.  Therefore, $\beta m=10$ is such kind of transition temperature.

}

\subsection{Comparison with general temperatures}\label{sec4.2}

For the case of $\beta m\sim 1$, the Eq.\eqref{1.5} for the low temperature limit in \cite{Camargo:2022rnt} is also not applicable. Only the approximation in Eq.\eqref{3.10} is feasible in this case. 

\begin{figure}[htb]
	\centering
	\subfigure[Krylov complexity for { $\beta m=1$}.]{\includegraphics[width=0.45\linewidth]{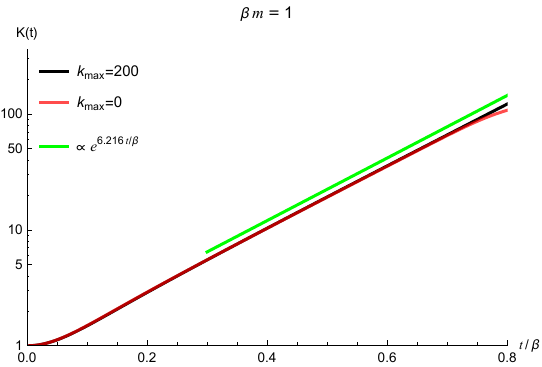}}~~
	\subfigure[Krylov entropy for { $\beta m=1$}.]{\includegraphics[width=0.45\linewidth]{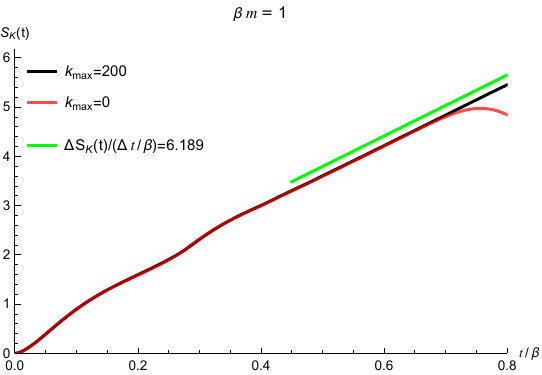}}
	\subfigure[Krylov complexity for { $\beta m=0.1$}.]{\includegraphics[width=0.45\linewidth]{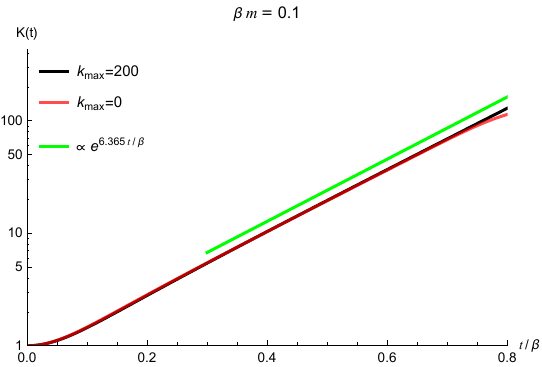}}~~
	\subfigure[Krylov entropy for { $\beta m=0.1$}.]{\includegraphics[width=0.45\linewidth]{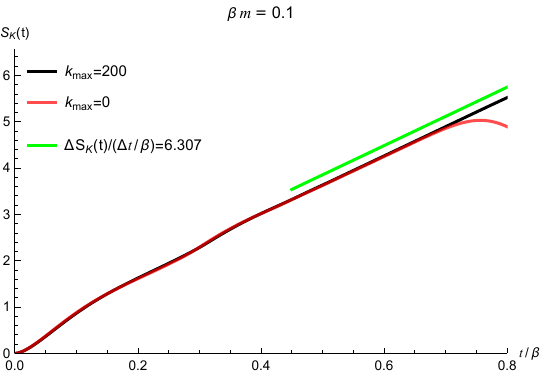}}
	\subfigure[Krylov complexity for { $\beta m=50$}.]{\includegraphics[width=0.45\linewidth]{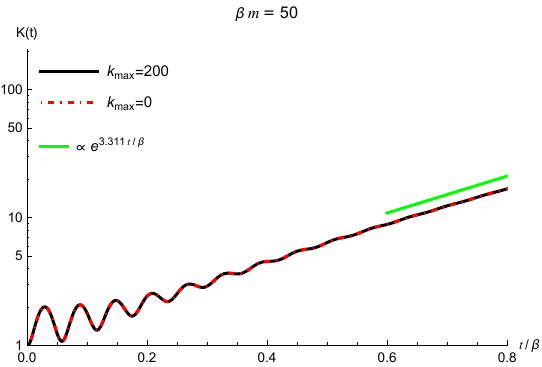}}~~
	\subfigure[Krylov entropy for { $\beta m=50$}.]{\includegraphics[width=0.45\linewidth]{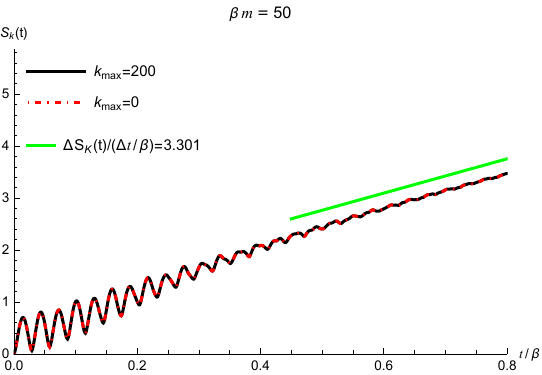}}
	\caption{Time evolutions of Krylov complexity $K(t)$ and Krylov entropy $S_K(t)$ { for various values of $\beta m$}. The panels (a), (c) and (e) for Krylov complexity are plotted in logarithmic scales. For all of the plots, the black lines are for $k_{\text{max}}=200$ corresponding to Eq.\eqref{3.10}, while the red (dashed) lines are for $k_{\text{max}}=0$ corresponding to Eq.\eqref{1.5}. The green lines are the reference lines for the scalings at late time.}
	\label{fig:kt11andst11}
\end{figure}

In Figure \ref{fig:kt11andst11}, we show the time evolution of the Krylov complexity $K(t)$ (in left column) and Krylov entropy $S_K(t)$ (in right column) for { different values of $\beta m$}. In these plots, the black curves are for $k_{\rm max}=200$ corresponding to Eq.\eqref{3.10} while the red (dashed) curves are for $k_{\rm max}=0$ corresponding to Eq.\eqref{1.5}. From the panels in the first row, i.e. { $\beta m=1$}, we can see that the red lines will deviate from the black lines at late time. This reflects that the approximations in Eq.\eqref{1.5} is not accurate as the approximations in Eq.\eqref{3.10} at late time, which confirms our assertion that the approximation in Eq.\eqref{1.5} is only applicable in low temperature limit, i.e., $\beta m\gg1$. The green lines in the first row are the reference lines for $K(t)$ and $S_K(t)$, respectively. In the panel (a) of Figure \ref{fig:kt11andst11} we can see that the exponential growth rate of the Krylov complexity is roughly $\lambda_K\approx 6.216$ which is close to $2\pi/\beta$. This exponential scaling is consistent with our discussions in the preceding subsection. In the panel (b) of Figure \ref{fig:kt11andst11}, the linear scaling of the Krylov entropy is roughly { $\Delta S_K(t)/(\Delta t/\beta)\approx6.189$}, which is consistent with the claims that $S_K(t)\sim\log K(t)$ at late time as we discussed in the preceding subsection. The small differences between the two scalings, i.e., $6.216$ and $6.189$ are due to the fittings of the two quantities at a finite time. It is expected that much longer time fitting will reduce this difference.   

In the second row of Figure \ref{fig:kt11andst11} we compare the Krylov complexity and Krylov entropy with distinct $k_{\rm max}$ (i.e., $k_{\rm max}=200$ and $k_{\rm max}=0$) in high temperature limit ($\beta m=0.1\ll1$). Again, the deviations between the red lines ($k_{\rm max}=0$) and black lines ($k_{\rm max}=200$) tell us that in the high temperature limit, the approximation in Eq.\eqref{1.5} is not valid. The late time scalings in $K(t)$ and $S_K(t)$ (blue reference lines) are also close to $2\pi/\beta$ and consistent with the preceding discussions. 

However, for $\beta m\gg1$, both the approximations in Eqs.\eqref{1.5} and \eqref{3.10} are valid, see for instance the last row in Figure \ref{fig:kt11andst11}, where we have set { $\beta m=50$} for the computation of Krylov complexity and Krylov entropy. In the time region we have considered, both of $k_{\rm max}=200$ and $k_{\rm max}=0$ collapse together and cannot be distinguished from each other. This is different from the cases with $\beta m=1$ and $\beta m=0.1$ in the first two rows of Figure \ref{fig:kt11andst11}. Interestingly, the late time scalings for both Krylov quantities are much smaller than $2\pi/\beta$, i.e., $\lambda_K<2\pi/\beta$ in the low temperature limit $\beta m\gg1$. Compared to high temperature limit, we may conjecture that the exponential growth rate of the Krylov complexity has a bound $\lambda_K\leq2\pi/\beta$, which is consistent with the discussions in \cite{Camargo:2022rnt}.

Another interesting thing is that in the low temperature limit $\beta m\gg1$, there are oscillations in $K(t)$ in the short time region. The period is roughly $\pi/m$. However, in the opposite limit, i.e., $\beta m\lesssim1$, there are no such oscillations in $K(t)$.  These phenomena are consistent with the observations in \cite{Camargo:2022rnt,Dymarsky:2021bjq}.

\section{Conclusions}\label{sec5}
In this paper, we have performed a detailed study of the Krylov complexity and Krylov entropy for a free massive scalar field in five-dimensional spacetime under various temperature conditions.  Previously, studies on the Krylov complexity of free scalar fields only considers the low-temperature limit, i.e., $\beta m\gg 1$, since in that case the Wightman power spectrum can be easily approximated as in Eq.\eqref{1.5}. In the general temperatures, i.e., general values of $\beta m$, we found a new expression for $f^{W}(\omega)$ in the form of an infinite series in Eq.\eqref{3.2}. We further obtained a good approximation of $f^{W}(\omega)$ by truncating the infinite summations to a finite summation, which allowed us to study the Krylov complexity and Krylov entropy in the regimes where $\beta m$ does not necessarily subject to $\beta m\gg1$. 


Then we examined the performance of the Lanczos coefficients $b_n$, which exhibited the `staggering' behavior by grouping $b_n$ into odd and even $n$ families. As usual, $b_{n}$ can be linearly fitted separately for odd and even $n$ for large $n$. The linear coefficients $\alpha$ satisfies the relation $\alpha\approx\pi/\beta$ which is consistent with the previous findings in the universal operator growth hypothesis.   We further observed that the difference $\gamma_{\text{odd}}-\gamma_{\text{even}}$ is linearly proportional to the mass $m$ for larger $m$, however, this proportionality breaks down for small $m$. This phenomenon was not explored in the existing literatures and we argued that the breakdown of the linear proportionality may come from the high temperature limit. 

Next, we investigated the cases for general temperatures with the general values of $\beta m$. In the late time, we found that the Krylov complexity exhibited an exponential growth $K(t)\sim e^{\lambda_K t}$ where the exponent satisfies the bound $\lambda_K\leq 2\pi/\beta$. The Krylov entropy is linearly proportional to $t$ in the late time and satisfying the relation $S(t)\sim \log K(t)$. In the high temperature limit $\beta m\ll 1$, the Krylov complexity exhibits behaviors remarkably similar to those of the conformal field theory where $\beta m=0$. Moreover, it was observed that in the high-temperature limit the Krylov complexity would monotonically increase, while in the low-temperature limit it would oscillate. We speculated that this difference was due to the staggering of $b_n$ as well as the function $\varphi_0(t)$. Our speculation was consistent with the numerical results. Subsequently, we numerically found that there existed a plateau in the Krylov complexity, which indicated that the temperature at this point was the transition temperature. Lower or higher temperatures would destroy this plateau, leading to oscillations or monotonic increasing of the Krylov complexity. Therefore, the temperature at which this plateau appears can be considered a ‘critical temperature’.


We also checked that the finite truncations $k_{\rm max}=200$ in our paper is accurate enough to get the correct results by comparing it to the previous studies with $k_{\rm max}=0$. Therefore, the schemes in our paper to study the Krylov complexity and Krylov entropy with general temperatures is reliable.



In conclusion, our study provides a comprehensive analysis of the Krylov complexity and Krylov entropy in a five-dimensional scalar field theory under varying temperature conditions. These findings enhance our understanding of operator growth and complexity in quantum field theories, particularly in relation to thermal effects. Future work may extend these results to other types of field theories and different spacetime dimensions, by exploring the rich structures of Krylov complexity in diverse physical settings.

\acknowledgments
This work was partially supported by the National Natural Science Foundation of China (Grants No.12175008).

\appendix

\section{Krylov complexity and Krylov entropy in symmetry breaking phase}\label{secssb}
In this appendix, we will study the Krylov complexity and Krylov entropy in the symmetry breaking phase by adopting a model of a five-dimensional real Higgs scalar field theory. The steps will be similar to those in the main context except that the mass $m$ will be changed to $M(\varphi)$ which will be shown later.  

Assume $\phi$ is a Higgs scalar field which can be described by the following Lagrangian,
\begin{equation}\label{A1}
	L_{\rm Higgs}=\frac{1}{2}\partial_{\mu}\phi\partial^{\mu}\phi-\frac{1}{2}m^{2}\phi^{2}-\lambda\phi^{4},
\end{equation}
in which $m^2<0$ and $\lambda>0$ in order to render the potential 
\be\label{higgspotential} U(\phi)=\frac{1}{2}m^{2}\phi^{2}+\lambda\phi^{4}\ee
behaves like a double well potential. According to the Appendix \ref{A}, we can write $\phi(X)$ as \eqref{A9}
\begin{equation}
	\phi(X)=\varphi+\sigma(X),
\end{equation}
where $\varphi$ is the classical value corresponding to the minimal point of the potential while $\sigma(X)$ is a new dynamical field. $X$ is the Euclidean coordinates of the spacetime. In the thermal field theory, the propagator of the $\sigma$ field is \eqref{A21}\footnote{For more details, see Appendix \ref{A}.}
\begin{equation}
	\mathcal{D}_{\sigma}(K)=\frac{1}{\omega_{n}^{2}+\mathbf{k}^{2}+M^{2}(\varphi)},
\end{equation}
where \be\label{Msq} M^{2}(\varphi)=m^{2}+12\lambda \varphi^{2}.\ee After quantum corrections in $d=5$ Higgs field theory, $\varphi$ here will take the value as \eqref{A43}
\begin{equation}
	\varphi_{\rm Q.C.}=\bar{\varphi}=\pm\frac{1}{2\pi}\sqrt{\frac{-m^{2}\pi^{2}-3\lambda T^{3}\zeta(3)}{\lambda}}.
\end{equation}

In computing the Krylov complexity the key point is to get the Wightman power spectrum $f^W(\omega)$ in Eq.\eqref{1.2}. However, $f^W(\omega)$ is calculated from the propagators in the field theory \cite{Camargo:2022rnt}. Compared to the free field theory in \cite{Camargo:2022rnt}, we find that the difference of the propagators between Higgs field theory and free field theory is the mass. Therefore, when calculating the Krylov complexity and Krylov entropy in the symmetry breaking phase, the only thing we need to do is to substitute $m$ in free field theory in the main context with the $M(\varphi)$ in Eq.\eqref{Msq}. From this respect, we will not show the figures of Krylov complexity and Krylov entropy further in the symmetry breaking phase. They are expected to behave similarly to those in the main context by just regarding $m$ as $M(\varphi)$.
	
\section{Recapitulation of Higgs Scalar Field Theory}\label{A}
In this appendix, we will compute the propagators and mass shift in the symmetry breaking phase of a real Higgs scalar field theory. 
In the formula of the Lagrangian Eq.\eqref{A1}, we adopt the metric signature as $(+,-,-,\cdots)$. One can check that the Lagrangian is invariant under $Z_{2}$ symmetry. 
The Higgs potential Eq.\eqref{higgspotential} has two global minima located at
\begin{equation}
	\phi=\pm v=\pm\sqrt{\frac{-m^{2}}{4\lambda}}.
\end{equation}
If we choose the vacuum expectation value $\expval{\phi}=v$, we can write $\phi(X)$\footnote{$\phi(X):=\phi(\tau,\bf{x})$ and we have used the imaginary-time formalism. In this formalism $t$ is replaced by $-i\tau$ and $\bf{x}$ represents the spatial coordinates.} as 
\begin{equation}
	\phi(X)=v+\sigma(X).
\end{equation}
The Lagrangian \eqref{A1} can be rewritten as 
\begin{equation}\label{A5}
	L_{\rm Higgs}=\frac{1}{2}\partial_{\mu}\sigma\partial^{\mu}\sigma-\frac{1}{2}(m^{2}+12\lambda v^{2})\sigma^{2}-4\lambda v\sigma^{3}-\lambda\sigma^{4}-U(v),
\end{equation}
where the tree-level potential reads
\begin{equation}
		U(v)=\frac{1}{2}m^{2}v^{2}+\lambda v^{4}.
\end{equation}
Therefore, Eq.\eqref{A5} describes a scalar field $\sigma$ with mass $\sqrt{m^{2}+12\lambda v^{2}}$ and it is not symmetric under the transformation $\sigma\leftrightarrow-\sigma$. The previous $Z_2$ symmetry is spontaneously broken. 

On the other hand, we can perform Fourier transformation for $\phi(X)$
\begin{equation}
	\phi(X)=\sqrt{\frac{\beta}{V}}\sum_{ K}e^{-iK\cdot X}\phi_{K},\qquad \sum_{K}\equiv\sum_{n=-\infty}^{\infty}\sum_{\mathbf{k}},\qquad -K\cdot X\equiv \omega_{n}\tau+\mathbf{k}\cdot \mathbf{x},
\end{equation} 
where $\omega_{n}\equiv 2\pi n/\beta ~(n\in\mathbb{Z})$ are called Matsubara frequency and $K^{\mu}=(-i\omega_{n},\mathbf{k}),X^{\mu}=(-i\tau,\mathbf{x})$. We can write down the partition function following the conventions in \cite{Kapusta:2006pm} that
\begin{equation}
	\mathcal{Z}=\int_{-\infty}^{\infty}d\phi_{0}\prod_{K\neq0}\int_{-\infty}^{\infty}d\Re\phi_{K}\int_{-\infty}^{\infty}d\Im\phi_{K}\exp(S[\phi]),
\end{equation}
in which $S[\phi]=\int_{0}^{\beta}d\tau d^{3}\mathbf{x}L_{\rm Higgs}$. By extracting out the zero mode of $\phi(X)$, we get
\begin{equation}\label{A9}
	\phi(X)=\sqrt{\frac{\beta}{V}}\sum_{K}\phi_{K}e^{-iK\cdot X}=\sqrt{\frac{\beta}{V}}\phi_{0}+\sqrt{\frac{\beta}{V}}\sum_{K\neq0}\phi_{K}e^{-iK\cdot X}=\varphi+\sigma(X),
\end{equation}
where $\varphi=v= \sqrt{\frac{\beta}{V}}\phi_{0}$ is independent of $X$ and $\sigma(X)\equiv \sum_{K\neq 0}e^{-iK\cdot X}\phi_{K}=\sum_{K\neq 0}e^{-iK\cdot X}\sigma_{K}$. 

Then we can define the effective potential as
\begin{equation}\label{A10}
	\mathcal{V}_{\text{eff}}(\varphi)=-\frac{T}{V}\ln(\prod_{K\neq0}\int_{-\infty}^{\infty}d\Re\sigma_{K}\int_{-\infty}^{\infty}d\Im \sigma_{K}\exp(S[\varphi+\sigma]))=-\frac{T}{V}\ln(\int\mathcal{D}\sigma e^{S[\varphi+\sigma]}),
\end{equation}
where $\int\mathcal{D}\sigma\equiv\prod_{K\neq0}\int_{-\infty}^{\infty}d\Re\sigma_{K}\int_{-\infty}^{\infty}d\Im \sigma_{K}$ and $S[\varphi+\sigma]=S[\phi]$. Then we have 
\begin{equation}
	\mathcal{Z}=\sqrt{VT}\int_{-\infty}^{\infty}d\varphi\exp(-\frac{V}{T}\mathcal{V}_{\text{eff}}(\varphi)).
\end{equation}
Without loss of generality, we assume that the minimum of effective potential is located at $\varphi=\bar{\varphi}>0$. In the thermodynamic limit $ V\rightarrow \infty$, the partition function can be written nearby $\bar\varphi$ as
\begin{equation}
		\mathcal{Z}=\exp(-\frac{V}{T}\mathcal{V}_{\text{eff}}(\bar \varphi))\sqrt{\frac{2\pi T^{2}}{\mathcal{V}^{\prime\prime}_{\text{eff}}(\bar{\varphi})}}.
\end{equation} 

Next, we need to compute $\bar{\varphi}$. To this end, we need to calculate the effective potential \eqref{A10} first. Note that $S[\varphi+\sigma]$ is actually the spacetime integral of the Lagrangian \eqref{A5} with $v=\varphi$. For convenience, we rewrite the Lagrangian as,
\begin{equation}
	\mathcal{L}[\sigma,\varphi]=\frac{1}{2}\partial_{\mu}\sigma\partial^{\mu}\sigma-\frac{1}{2}M^{2}(\varphi)\sigma^{2}-4\lambda\varphi\sigma^{3}-\lambda\sigma^{4}-U(\varphi),
\end{equation}
where 
\begin{gather}
	M^{2}(\varphi)=m^{2}+12\lambda\varphi^{2},\\
	U(\varphi)=\frac{1}{2}m^{2}\varphi^{2}+\lambda \varphi^{4}.
\end{gather}
Treating $\sigma$ as a dynamical field, the Lagrangian can be divided into a free part $\mathcal{L}_{0}$ and an interacting part $\mathcal{L}_{I}$ as,
\begin{equation}
	\mathcal{L}[\sigma,\varphi]=\mathcal{L}_{0}+\mathcal{L}_{I}-U(\varphi),
\end{equation}
where
\begin{equation}
	\mathcal{L}_{0}=\frac{1}{2}\partial_{\mu}\sigma\partial^{\mu}\sigma-\frac{1}{2}M^{2}(\varphi)\sigma^{2},\qquad \mathcal{L}_{I}=-4\lambda\varphi\sigma^{3}-\lambda \sigma^{4}.
\end{equation}
Since $\varphi$ is independent of $X$ and $\sigma$, the effective potential \eqref{A10} can be decomposed into (notice that $T=\beta^{-1}$)
\begin{equation}
	\mathcal{V}_{\text{eff}}(\varphi)=U(\varphi)-\frac{T}{V}\ln\left (\int\mathcal{D}\sigma e^{S_{0}[\sigma]+S_{I}[\sigma]}\right ),
\end{equation}
in which $S_{0}[\sigma]=\int_{0}^{\beta}d\tau d^{3}\mathbf{x}\mathcal{L}_0$ and $S_{I}[\sigma]=\int_{0}^{\beta}d\tau d^{3}\mathbf{x}\mathcal{L}_I$. 

In perturbation theory, the effective potential can be expanded in terms of $\hbar$ as (we have set $\hbar\equiv1$) 
\begin{equation}
	\mathcal{V}_{\text{eff}}(\varphi)=\mathcal{V}^{(0)}_{\text{eff}}(\varphi)+\mathcal{V}^{(1)}_{\text{eff}}(\varphi)+\mathcal{V}_{\text{eff}}^{(2)}+\cdots,
\end{equation}
where the superscript index $i$ in $\mathcal{V}^{(i)}_{\text{eff}}$ indicates the perturbations in the $i$-th power in $\hbar$.  Obviously, $\mathcal{V}_{\text{eff}}^{(0)}=U(\varphi)$ and 
\begin{gather}
	\mathcal{V}_{\text{eff}}^{(1)}(\varphi)=-\frac{T}{V}\ln\left (\int\mathcal{D}\sigma e^{S_{0}[\sigma]}\right )=\frac{1}{2}\frac{T}{V}\sum_{K}\ln(\frac{\mathcal{D}_{\sigma}^{-1}(K)}{T^{2}}),\\
	 \mathcal{D}_{\sigma}(K)=\frac{1}{\omega_{n}^{2}+\mathbf{k}^{2}+M^{2}(\varphi)}.\label{A21}
\end{gather}
The value of $\varphi$ needs to be determined by the minimum point of the effective potential. We will only consider zeroth order and first order terms of the effective potential, i.e., $\mathcal{V}_{\text{eff}}^{(0)}(\varphi)$ and $\mathcal{V}_{\text{eff}}^{(1)}(\varphi)$. Assuming the spacetime is $d$-dimensional, then following the steps in thermal field theory \cite{Laine:2016hma}, we get
\begin{gather}
	\mathcal{V}^{(0)}_{\text{eff}}(\varphi)=\frac{1}{2}m^{2}\varphi^{2}+\lambda\varphi^{4},\\
	\mathcal{V}^{(1)}_{\text{eff}}=\int\frac{d^{d-1}\mathbf{k}}{(2\pi)^{d-1}}\left [\frac{E_{\mathbf{k}}}{2}+T\ln(1-e^{-E_{\mathbf{k}}/T})\right ],
\end{gather}
where $E_{\mathbf{k}}=\sqrt{\mathbf{k}^{2}+M^{2}(\varphi)}$. We can recombine these two equations into a $T$-dependent part and a $T$-independent part
\begin{gather}
	\mathcal{V}_{0}(\varphi)=\frac{1}{2}m^{2}\varphi^{2}+\lambda\varphi^{4}+\int\frac{d^{d-1}\mathbf{k}}{(2\pi)^{d-1}}\frac{E_{\mathbf{k}}}{2},\label{A24}\\
	\mathcal{V}_{T}(\varphi)=T\int\frac{d^{d-1}\mathbf{k}}{(2\pi)^{d-1}}\ln(1-e^{-E_{\mathbf{k}}/T}).\label{A25}
\end{gather}
Using the identity \cite{Dolan:1973qd}
\begin{equation}
	E_{\mathbf{k}}=\int_{-\infty}^{\infty}\frac{dk_{0}}{2\pi i}\ln(-k_{0}^{2}+E_{\mathbf{k}}^{2}-i\epsilon),
\end{equation}
the integration in \eqref{A24} can be rewritten as 
\begin{equation}
	J(M)\equiv\int\frac{d^{d-1}\mathbf{k}}{(2\pi)^{d-1}}\frac{E_{\mathbf{k}}}{2}=-\frac{i}{2}\int\frac{d^{d}K}{(2\pi)^{d}}\ln(-K^{2}+M^{2}-i\epsilon),
\end{equation}
where we have set $K^{\mu}=(k^{0},\mathbf{k})$ and $M=M(\varphi)$. We can rotate the $k_{0}$-integral onto the imaginary axis,let $k_{0}=ik_{4}$, without crossing the branching cuts of the logarithm. With Euclidean $d$-momentum variable $K_{E}\equiv (k_{d},\mathbf{k})$ we have
\begin{equation}
	J(M)\equiv\frac{1}{2}\int\frac{d^{d}K_{E}}{(2\pi)^{d}}\ln(K^{2}_{E}+M^{2}).
\end{equation}
Using the following identities \cite{Peskin:1995ev}
\begin{equation}
	\int\frac{d^{d}K_{E}}{(2\pi)^{d}}=\int\frac{d\Omega_{d}}{(2\pi)^{d}}\int_{0}^{\infty}dK_{E}K^{d-1}_{E},\qquad \int d\Omega_{d}=\frac{2\pi^{d/2}}{\Gamma(d/2)},
\end{equation}
we have
\begin{equation}
	J(M)=\frac{1}{2}\frac{2\pi^{d/2}}{\Gamma(d/2)(2\pi)^{d}}\int_{0}^{\infty}dxx^{d-1}\ln(x^{2}+M^{2}).
\end{equation}
However, this integral is divergent. Therefore, we need to introduce a truncation $\Lambda$ such that
\begin{equation}
	\begin{aligned}
		J(M)&=\frac{1}{2}\frac{2\pi^{d/2}}{\Gamma(d/2)(2\pi)^{d}}\int_{0}^{\Lambda}dxx^{d-1}\ln(x^{2}+M^{2})\\
		&=\frac{1}{2d}\frac{2\pi^{d/2}}{\Gamma(d/2)(2\pi)^{d}}\left [\Lambda^{d}\ln(\Lambda^{2}+M^{2})-\int_{0}^{\Lambda}\frac{2x\cdot x^{d}}{x^{2}+M^{2}}dx\right ]\\
		&=\frac{1}{2d}\frac{2\pi^{d/2}}{\Gamma(d/2)(2\pi)^{d}}\left \{\Lambda^{d}\ln(\Lambda^{2}+M^{2})-\frac{2\Lambda^{2+d}}{(2+d)M^{2}}{}_{2}F_{1}\left [1,\frac{2+d}{2},\frac{4+d}{2},-\frac{\Lambda^{2}}{M^{2}}\right ]\right \},
	\end{aligned}
\end{equation}
where ${}_{2}F_{1}$ is the hypergeometric function. In the following, we will compute $J(M)$ with the dimensions with $d=4$ and $d=5$. 

\subsection{$d=4$}
 If $d=4$, we can expand $J(M)$ in terms of $1/\Lambda$
\begin{equation}
	J(M)=\left \{\frac{1}{32\pi^{2}}\left [M^{2}\Lambda^{2}-\frac{M^{4}}{2}\left (\ln\frac{\Lambda^{2}}{M^{2}}+\frac{1}{2}\right )\right ]\right \}+\frac{\Lambda^{4}}{64\pi^{2}}\left [\ln \Lambda^{2}-\frac{1}{2}\right ]+O(1/\Lambda^{2}),
\end{equation}
in which the second term is divergence and independent of $\varphi$, therefore, we can directly drop it. Let 
\begin{equation}
	\ln\frac{\Lambda^{2}}{M^{2}}=\ln\frac{\Lambda^{2}}{\mu^{2}}+\ln\frac{\mu^{2}}{M^{2}},\qquad {\rm with}~~M^{2}=m^{2}+12\lambda \varphi^{2},
\end{equation}
where $\mu$ is an arbitrary energy scale. Then we have
\begin{equation}
	\begin{aligned}
		J(M)=&\left [\frac{\Lambda^{2}m^{2}}{32\pi^{2}}-\frac{m^{4}}{64\pi^{2}}\ln\frac{\Lambda^{2}}{\mu^{2}}+\varphi^{2}\left (\frac{3\Lambda^{2}}{8\pi^{2}}\lambda-\frac{3m^{2}}{8\pi^{2}}\lambda\ln\frac{\Lambda^{2}}{\mu^{2}}\right )-\varphi^{4}\frac{9\lambda^{2}}{4\pi^{2}}\ln\frac{\Lambda^{2}}{\mu^{2}}\right ]\\
		&-\frac{m^{4}}{128\pi^{2}}-\frac{m^{4}}{64\pi^{2}}\ln\frac{\mu^{2}}{m^{2}+12\varphi^{2}\lambda}-\varphi^{2}\left (\frac{3m^{2}}{16\pi^{2}}\lambda+\frac{3m^{2}\lambda}{8\pi^{2}}\ln\frac{\mu^{2}}{m^{2}+12\varphi^{2}\lambda}\right )\\
		&-\varphi^{4}\left (\frac{9\lambda^{2}}{8\pi^{2}}+\frac{9\lambda^{2}}{4\pi^{2}}\ln\frac{\mu^{2}}{m^{2}+12\varphi^{2}\lambda}\right ).
	\end{aligned}
\end{equation}
The part in the square brackets $[\dots]$ is divergent and can be directly discarded. In addition, $-\frac{m^4}{128\pi^2}$ is independent of $\varphi$,  thus can also be discarded without affecting our results. Therefore, the part of the effective potential \eqref{A24} that does not depend on the temperature is
\begin{equation}
	\mathcal{V}_{0}(\varphi)=\left (1-\frac{3\lambda}{8\pi^{2}}\right )\frac{m^{2}}{2}\varphi^{2}+\left (1-\frac{9\lambda}{8\pi^{2}}\right )\lambda\varphi^{4}+\frac{1}{64\pi^{2}}\left (m^{2}+12\lambda\varphi^{2}\right )^{2}\ln\left (\frac{m^{2}+12\lambda\varphi^{2}}{\mu^{2}}\right ).
\end{equation}
Assume that the coupling constant $\lambda \ll 1$,  then \cite{Laine:2016hma}
\begin{equation}\label{A34}
	\mathcal{V}_{0}(\varphi)=\frac{m^{2}}{2}\varphi^{2}+\lambda\varphi^{4},
\end{equation}
where we have discarded all terms that do not depend on $\varphi$, as such terms will not affect our results. Now consider the part of the effective potential that depends on temperature \eqref{A25}
\begin{equation}
	\begin{aligned}
		\mathcal{V}_{T}(\varphi)&=T\int\frac{d^{3}\mathbf{k}}{(2\pi)^{3}}\ln(1-e^{-E_{\mathbf{k}}/T})\\
		&=4\pi T^{4}\int_{0}^{\infty}\frac{dx}{(2\pi)^{3}}x^{2}\ln(1-e^{-\sqrt{x^{2}+\alpha^{2}}}),
	\end{aligned}
\end{equation}
where $\alpha=M/T$. In the high temperature limit $\alpha\ll 1$,
\begin{equation}
	\ln(1-e^{-\sqrt{x^{2}+\alpha^{2}}})\approx\ln(1-e^{-x})+\frac{1}{e^{x}-1}\frac{\alpha^{2}}{2x}.
\end{equation}
Then we get
\begin{equation}\label{A37}
	\mathcal{V}_{T}(\varphi)=-\frac{\pi^{2}}{90}T^{4}+\frac{T^{2}}{24}(m^{2}+12\varphi^{2}\lambda),
\end{equation}
in which the terms independent of $\varphi$ can be discarded again. Therefore, combining Eqs.\eqref{A34} and \eqref{A37} the effective potential under weak coupling $\lambda\ll1$ and high temperature $\alpha\ll1$ is \cite{Laine:2016hma}
\begin{equation}
	\mathcal{V}_{\text{eff}}(\varphi)=\frac{1}{2}(m^{2}+\lambda T^{2})\varphi^{2}+\lambda\varphi^{4}.
\end{equation}
The minimum point of the effective potential is located at
\begin{equation}
	\bar{\varphi}=\pm\frac{1}{2}\sqrt{\frac{-m^{2}-\lambda T^{2}}{\lambda}}.
\end{equation}

\subsection{$d=5$}
Following the previous operations, we can obtain the effective potential with $d=5$. The specific process will not be repeated here. The part of the effective potential that does not depend on temperature is
\begin{equation}
	\begin{aligned}
		\mathcal{V}_{0}^{(5)}(\varphi)=&\left (1+\frac{2\lambda \sqrt{m^{2}+12\lambda\varphi^{2}}}{5\pi^{2}}\right )\frac{m^{2}}{2}\varphi^{2}+\left (1+\frac{6\lambda \sqrt{m^{2}+12\lambda\varphi^{2}}}{5\pi^{2}}\right )\lambda \varphi^{4}\\
		&+\frac{m^{4}\sqrt{m^{2}+12\lambda\varphi^{2}}}{120\pi^{2}}.
	\end{aligned}
\end{equation}
The part of the effective potential that depends on temperature is
\begin{equation}
	\mathcal{V}_{T}^{(5)}(\varphi)=\frac{3\lambda T^{3}\varphi^{2}\zeta(3)}{2\pi^{2}},
\end{equation}
where $\zeta$ is the Riemann-zeta function. In the limit of high temperature and weak coupling, the effective potential is
\begin{equation}
	\mathcal{V}^{(5)}_{\text{eff}}(\varphi)=\frac{m^{2}}{2}\varphi^{2}+\lambda\varphi^{4}+\frac{3\lambda T^{3}\varphi^{2}\zeta(3)}{2\pi^{2}}=\frac{1}{2}\left (m^{2}+\frac{3\lambda T^{3}\zeta(3)}{\pi^{2}}\right )\varphi^{2}+\lambda\varphi^{4}.
\end{equation}
The minimum point of the effective potential is located at
\begin{equation}\label{A43}
	\bar{\varphi}=\pm\frac{1}{2\pi}\sqrt{\frac{-m^{2}\pi^{2}-3\lambda T^{3}\zeta(3)}{\lambda}}.
\end{equation}






\bibliographystyle{jhep}
\bibliography{references.bib}
\end{document}